\documentclass[
  reprint,
  superscriptaddress,
  amsmath,amssymb,
  aps,
  prb,
  floatfix
]{revtex4-2}

\usepackage{graphicx}
\usepackage{dcolumn}
\usepackage{bm}
\usepackage{epstopdf}
 \usepackage[dvipsnames]{pstricks}
 \usepackage{epsfig}
 \usepackage{pst-grad} 
 \usepackage{pst-plot} 
 \usepackage[space]{grffile} 
 \usepackage{etoolbox} 
 \makeatletter 
 \patchcmd\Gread@eps{\@inputcheck#1 }{\@inputcheck"#1"\relax}{}{}
 \makeatother
\usepackage{booktabs}
\usepackage{boxhandler}
\usepackage{amsmath}
\usepackage{bbold}
\usepackage{hyperref}
\usepackage{natbib}
\usepackage[whole]{bxcjkjatype}
\usepackage{tikz}
\usepackage{mathtools}
\usepackage[normalem]{ulem}

\newcommand{\stkout}[1]{\ifmmode\text{\sout{\ensuremath{#1}}}\else\sout{#1}\fi}

\usetikzlibrary{arrows.meta}

\usepackage[caption=false]{subfig}
\usepackage{floatrow}
\usepackage[utf8]{inputenc}
\usepackage{braket}
\usepackage{multirow}
\usepackage{enumitem}
\usepackage{pifont}
\usepackage{lmodern}

\renewcommand{\vec}[1]{\bm{#1}}

\hypersetup{
  colorlinks   = true,    
  urlcolor     = blue,    
  linkcolor    = blue,    
  citecolor    = blue      
}

\begin{document}

\title{Impact of leakage on the dynamics of a ST$_0$ qubit implemented in a Double Quantum Dot device}
\author{Javier Oliva del Moral}
\email{jolivam@unav.es}
\affiliation{Department of Basic Sciences, Tecnun - University of Navarra, 20018 San Sebastian, Spain.}
\affiliation{Donostia International Physics Center, 20018 San Sebastián, Spain.}
\author{Olatz Sanz Larrarte}
\affiliation{Department of Basic Sciences, Tecnun - University of Navarra, 20018 San Sebastian, Spain.}
\author{Reza Dastbasteh}
\affiliation{Department of Basic Sciences, Tecnun - University of Navarra, 20018 San Sebastian, Spain.}
\author{Josu Etxezarreta Martinez}
\affiliation{Department of Basic Sciences, Tecnun - University of Navarra, 20018 San Sebastian, Spain.}
\affiliation{Visiting Researcher at the Cavendish Laboratory, Department of Physics, University of Cambridge, Cambridge CB3 0HE, UK.}
\author{Rubén M. Otxoa}
\affiliation{Hitachi Cambridge Laboratory, J. J. Thomson Avenue, Cambridge CB3 0HE, United Kingdom.}

\begin{abstract}
Spin qubits in quantum dots are a promising technology for quantum computing due to their fast response time and long coherence times. An electromagnetic pulse is applied to the system for a specific duration to perform a desired rotation. To avoid decoherence, the amplitude and gate time must be highly accurate. In this work, we aim to study the impact of leakage during the gate time evolution of a spin qubit encoded in a double quantum dot device. We prove that, in the weak interaction regime, leakage introduces a shift in the phase of the time evolution operator, causing over- or under-rotations. Indeed, controlling the leakage terms is useful for adjusting the time needed to perform a quantum computation and increasing the coherence time of the readout process. This is crucial for running fault-tolerant algorithms and is beneficial for Quantum Error Mitigation techniques.
\end{abstract}

\maketitle

\section{Introduction}
\noindent Quantum Dots (QDs) hosting spin qubits are an innovative technology for building quantum computers leveraging the spin state of electrons confined in semiconductor nano-structures. These QDs are created by gate electrodes patterned on top of a semiconductor which induce a confining potential for electrons~\cite{QuantumDots}. The number of electrons can be controlled by manipulating the applied potential~\cite{CblockadeFujisawa_2003}. The spin of the electron serves as quantum bit unit and can be manipulated by applying external electric and magnetic fields~\cite{LossDivincenzo}. When an external magnetic field is applied perpendicular to the plane of the QD, the energy of the degenerate $\ket{\uparrow}$ and $\ket{\downarrow}$ states is split due to the Zeeman interaction so that they can be used as the computational basis where the qubit is encoded. Other alternatives exist for encoding a qubit using more than a single QD, including singlet-triplet spin qubits, proposed in~\cite{Levy} and firstly demonstrated experimentally in~\cite{T2noise}; and flip-flop qubits~\cite{Tosi2017}, among others~\cite{QuantumDots}. Key features of spin qubits in QDs are their rapid response time to external electric and magnetic fields resulting in fast quantum gates and long coherence times~\cite{LongCoherenceT}. As a result, they stand as a promising technology for developing quantum computers. However, several technological challenges related to control, scalability, and connectivity must be overcome to make such proposal a reality~\cite{Control,Scaling,connectivity,Vandersypen2017}. The control needed to perform qubit rotations depends on various factors such as: the encoding protocol, the interaction strength, the inherent properties of the used semiconductor and the dimensionality of the QD~\cite{QuantumDots}. Consequently, a crucial aspect of spin qubit manipulation is the fidelity achieved when performing qubit rotations by means of external electric and/or magnetic fields. This issue is of paramount importance as running quantum algorithms typically requires a large number of rotations or quantum gates, each of which must be highly precise to prevent coherent errors arising from under- or over-rotations~\cite{Cai2020}. Incoherent errors arising from the decoherence of the quantum information as result of its interaction with the environment are also an important issue and unavoidably leads to imprecise results~\cite{TonDecoders,QEM,ZNE}. To circumvent this problem, Quantum Error Mitigation (QEM) techniques~\cite{QEM}, such as zero-noise extrapolation (ZNE), aim to enhance the accuracy of quantum algorithms by post-processing several noisy measurements of expectation values of interest~\cite{ZNE,Comment}. These methods provide a promising strategy for harnessing useful outcomes from quantum computers in the noisy intermediate-scale quantum (NISQ) era until the development of fault-tolerant quantum processors~\cite{Preskill}. 

Fault-tolerant quantum computing refers to the paradigm of being able to reliably perform quantum computations even if its constituent elements are imperfect. The building block of such machines is the so-called quantum error correction (QEC). QEC codes use a large number of physical qubits ($n$) to encode a few logical qubits $k<n$, surface codes use many physical qubits to protect a single logical qubit $k = 1$. The QEC process requires many parity check measurements to detect and correct errors~\cite{QEC}. Consequently, the fidelity of the readout process is a critical metric for quantum hardware. Despite recent breakthrough advances in this field, such as the experimental demonstration on a superconducting qubit chip that surface codes can suppress the error rate when operating at sub-threshold error rates~\cite{google24}, it has not yet been proven that a universal fault-tolerant quantum computers can be constructed with current technology. The main challenges are related to qubit connectivity as well as the challenge of achieving a universal set of quantum gates for the logical qubits encoded on stabilizer codes~\cite{FaultTUnivCampbell2017}, reaching physical error rates below the threshold for certain technologies~\cite{gottesman2009introductionquantumerrorcorrection} and real-time decoders~\cite{TonDecoders}, among others. Furthermore, leakage has been shown to be extremely detrimental for QEC implemented in superconducting qubits~\cite{leakageqec1}. In the context of semiconductor spin qubits, many proposals are being considered to construct fault-tolerant quantum computers, using standard Loss-Divincenzo (LD) qubits, singlet-triplet qubits and hybrid approaches~\cite{spinQEC,SpinHEx,LDSTqec,2Narray}.

Therefore, it is essential to fabricate qubits with long coherence times and to implement high-fidelity gates, measurements and state preparations. In this paradigm, a key challenge related to gate implementation is designing accurate pulses to prevent over- or under-rotations (coherent errors) and optimize timing, since faster gates may result in lower incoherent errors~\cite{Mei-YaChen2022}. Furthermore, the systems in which qubits are encoded may not be strictly two-level systems, i.e. more energy levels than the computational subspace (where the qubit is encoded) may exist and be accessible. The presence of those other levels introduces extra transitions or pathways between qubit states, potentially altering their dynamics~\cite{leakage}. In this study, we investigate how these additional states affect the quantum time evolution operators when a constant pulse is applied over time. Interestingly, this corresponds to a scenario that can be tested experimentally with certain ease~\cite{PulseExperi}.

Throughout this work, we focus on singlet-triplet (ST$_0$) qubits encoded in a double quantum dot (DQD) device because they possess more than two accessible energy levels and transitions between the computational basis ($\ket{S}$ and $\ket{T_0}$)  and the states $\ket{T_+}$ and $\ket{T_-}$ by means of magnetic fields. Additionally, the atomic structure of the material~\cite{jock2022} and the specific geometry of the QDs~\cite{RevModPhysSiQ} could also provide access to additional energy levels. The theoretical approach we present in this work is generalizable to any source of leakage and to any qubit technology with more than two accessible energy levels. There are different works that focus on the effect of leakage in other technologies, such as superconducting qubits~\cite{LeakingQb,leakage}. Here, we study the impact of leakage on the unitary evolution operator of the ST$_0$ qubits. However, in a realistic scenario coherence times are strongly limited by the hyperfine interaction and the gate fidelity are limited by the charge noise, see for example~\cite{Zhang2025,sgn1-1t2d,Kawakami2014}. Therefore, we include a small analysis of the gate fidelities of these qubits in presence of noise and a transverse magnetic field.

The article is organized as follows: Section~\ref{Sec: Motivation} provides an introduction to the quantum time evolution operator and the physics of QD and DQD systems. Next, Section~\ref{Sec: Time Evolution} focuses on the unitary evolution of ST$_0$ qubits, where the logical states are encoded in the singlet and neutral triplet states of the DQD. Finally, we compute the fidelity of the evolution operators in the presence of decoherence. Through a representative analysis, this study provides a bound on the impact of leakage on gates fidelity. The study concludes with Section~\ref{Sec: Discussion}, which offers a discussion, and Section~\ref{Sec: Conclusion}, which presents the conclusions along with a brief discussion of quantum system time evolution in the presence of leakage.

\section{Theoretical background} \label{Sec: Motivation}

\begin{figure}[t!]
    \centering
    \subfloat[]{\includegraphics[clip,width=.9\columnwidth]{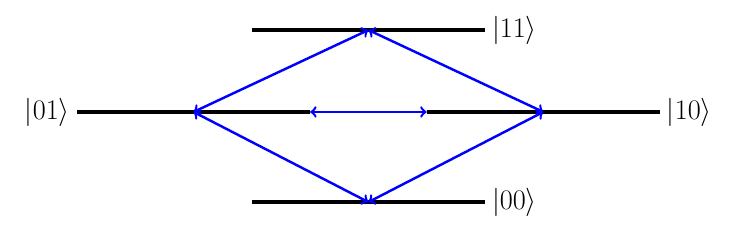}\label{fig:Two level}} \\
    \subfloat[]{\includegraphics[clip,width=.89\columnwidth]{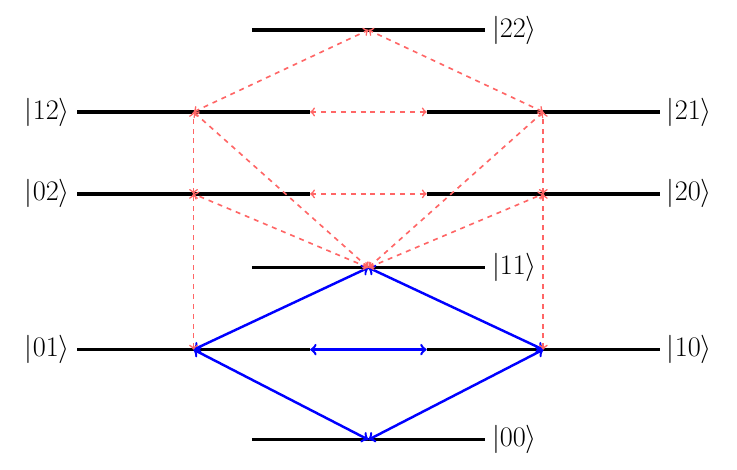}\label{fig:Three level}}

    \caption{Graphical representation of the possible states of (a) an ensemble of two TLSs (or qubits) and (b) an ensemble of two three level systems (or qutrits) with the transitions between the states (blue and red lines). The energy of each state is given by the respective diagonal term of the Hamiltonian, $\text{H}_0$, and the transitions are related to the off-diagonal terms of the Hamiltonian, $\text{H}_I$. Blue (solid) lines represent rotations of the qubit, i.e. in the SU(2) subspace, and red (dashed) lines represent leakage transitions to higher energy levels and between them.}
    
\end{figure}

The dynamics of a quantum system is described by its Hamiltonian, which defines the energies of the different possible states and the transitions among them. Is important to note that the number of paths from a given state to another one has an impact on the dynamics. For instance, in a two-level system (TLS), the transition from an initial state to a final state is uniquely defined by the direct transition, meaning there is only one possible path from $\ket{0}$ to $\ket{1}$. For an ensemble of two TLSs, the possible transitions are represented by blue lines in Fig. \ref{fig:Two level}. However, whenever more accessible levels with allowed transitions among them exist, the evolution between the states becomes more complex. A graphical representation of an ensemble of two quantum systems with three accessible levels, i.e. qutrits, and transitions among these levels is shown in Fig.~\ref{fig:Three level}. Blue (solid) lines represents transitions between the states which conforms the qubit and red (dashed) lines represents leakage transitions to higher energy levels. Note that we consider that qubits are encoded in those three level systems (depicted in blue in Fig.~\ref{fig:Three level}), but the other levels out of the computational subspace are still accessible (depicted in red in Fig.~\ref{fig:Three level}). The difference in how these two scenarios evolve over time is the main question we analyze in this paper, i.e. a strictly two-level system qubit versus a qubit encoded in a higher dimensional quantum state.

Generally, a qubit encoded on a two-level system is defined by a $2 \times 2$ Hamiltonian $\text{H}_0= \lambda \sigma_z$, where $\sigma_z$ is the Pauli operator in direction $\hat{z}$. To perform a rotation, defined as $R_{i}(\theta)$, of $\theta$ radians around $i\in\{\hat{x},\hat{y},\hat{z}\}$, we have to add an interaction term $H_I$ defined by the Pauli matrices $\sigma_x$, $\sigma_y$ or $\sigma_z$, respectively:
\begin{align} \label{Eq: BasicRot}
    R_i(\theta) = \exp{\left(-i \sigma_i \frac{\theta}{2}\right) },
\end{align}
where $\sigma_i\in \{\sigma_x,\sigma_y,\sigma_z\}$. Generally, this interaction term is enforced to the system by means of an external pulse applied to it. Importantly, since the applied pulses cannot have arbitrary shapes, we assume that rectangular shapes are used for enforcing these rotations, which are more feasible to be implemented experimentally~\cite{Kawakami2014,T2noise}.

While considering a qubit as a strict TLS is a useful abstraction, real quantum systems often include more possible energy levels that can be accessed. A qubit is encoded using two of those levels, defining a computational subspace. Ideally, transitions from the computational basis, defined by a $2 \times 2$ Hamiltonian, to higher energy states are considered to be prohibited. However, a direct consequence of the existence of those extra accessible levels is that the qubit leaves the computational basis $\ket{0}$ and $\ket{1}$. This effect is known as \textit{leakage}~\cite{leakage,leakageqec1,leakageqec2,LeakageErrors}. The existence of leakage makes the number of possible paths from an initial state to a final state to increase (as it is illustrated in Fig.~\ref{fig:Three level}), thus, changing the time evolution operator. This work focuses on how leakage breaks some symmetries of the system allowing new possible paths from an initial state to a final state. We employ the Dyson series expansion, which aligns with our symmetry-based interpretation of the problem and can be extended to time-dependent Hamiltonians. However, there are other methods to study these kind of systems, such as the Schrieffer–Wolff (SW) transformation~\cite{SW1} which, for example, has been used to study the readout process of ST$_0$ qubits in~\cite{ReadoutDecay}. In the following section, we present how qubits are encoded in a QD and in a DQD device, which has more than two accessible levels, implying that leakage affects the dynamics

\subsection{Spin Qubits in Quantum Dots} \label{Sec: Spin Qubits}

In the following, we present the most important concepts of spin qubits in QDs, while additional information can be found in the Appendix~\ref{SecA: Spin Qubits}. In addition, several excellent works address this topic, such as Refs.~\cite{Burkard2002,Harvey_2022,QDreview1}. We encourage interested readers to consult these references for more detailed information. 

Spin qubits are encoded in the spin state of an electron confined within a QD formed in a semiconductor~\cite{Burkard2002}. The electron spin state interacts with an external time-dependent magnetic field, $\vec{B}(t)$, through the Zeeman Hamiltonian, $\hat{H}_{\text{Zeeman}}(\vec{B}(t))$ and with another spin qubit via the Heisenberg exchange interaction, $\hat{H}_{\text{exc.}}( \text{J}_\varepsilon(t))$, where the exchange coupling strength $  \text{J}_\varepsilon(t)$ is tunable via the voltage detuning parameter $\varepsilon(t)$~\cite{QuantumDots}. We can perform single qubit gates by changing the external magnetic field using the Zeeman Hamiltonian defined as:

\begin{align} \label{Eq: Zeeman Hamiltonian}
    \text{H}_Z(t) =  g \mu_B \vec{B}(t) \cdot \vec{s},
\end{align}

where g is the gyromagnetic ratio of the electron, $\mu_B$ is the Bohr magneton, $\vec{B}$ is the external magnetic field and $\cdot$ refers to the dot product. $\vec{s}= \frac12 \vec{\sigma}$ is the spin operator vector, with $\vec{\sigma}=\{\sigma_x, \sigma_y, \sigma_z\}$ the vector of Pauli matrices. As a result, the rotation operator is defined in Eq.~4,  with a rotation angle $\theta(\lambda_x,\tau_{G_x})$, which depends on the amplitude of the interaction $\lambda_i = \frac12 g \mu_B B_i$ and the gate time $\tau_{G_i}$, $\forall i \in \{x,y,z\}$. Two qubit (entangling) gates can be implemented by means of both the Zeeman and the exchange interactions, the Hamiltonian is given by:

\begin{align} \label{Eq: 2qubit}
    \text{H}_C(t) &= \text{H}_\text{exc}(t) + \sum_{j=1,2} \text{H}_{Z,j}(t) \\
    &=   \text{J}_\text{exc}[\varepsilon(t)] \vec{s}_1 \cdot \vec{s}_2 +  \sum_{i=1,2} g \mu_B \vec{B}(r_i,t) \cdot \vec{s}_i,
\end{align}

where $\pm  \text{J}_\text{exc}[\varepsilon(t)]$ is the exchange coupling strength between them, which can be controlled by means of the \textit{voltage} detuning between their respective potentials $\varepsilon = V_{\text{Ext}_1}-V_{\text{Ext}_2}$, where $V_{\text{Ext}_i}$ is the potential applied to the i$^\text{th}$; and $\vec{B}_i(r_i,t)$ is the magnetic field in the position of the i$^{th}$ QD. Henceforth, we refer to the magnetic fields as $b_i = \frac12 g \mu_B B_i$ and $\delta b_i = \frac12 g \mu_B \delta B_i$; and the exchange coupling as $J\equiv \frac{1}{8} \text{J}_\text{exc}[\varepsilon]$ to simplify the notation. 

Quantum information can be also encoded in the singlet $\ket{S}$  and triplets $\ket{T_{0,\pm}}$ states of a DQD device with one electron bounded to each QD. The Hamiltonian of this quantum system in the  $\{\ket{S}$, $\ket{T_{0}}$, $\ket{T_+}$, $\ket{T_-}\}$ basis is given by:

\begin{widetext}
\begin{align} \label{Eq: Hamil6x6}
\text{H}&= \text{H}_\text{Z} + \text{H}_\text{exc}      =\begin{pmatrix}
          -\text{J} & \delta b_z & -\frac{1}{\sqrt{2}}  [\delta b_x+i\delta b_y] & \frac{1}{\sqrt{2}}  [\delta b_x-i\delta b_y] \\
          \delta b_{z} & \text{J}& \frac{1}{\sqrt{2}}  [b_x+ib_y] & \frac{1}{\sqrt{2}}  [b_x- i  b_y] \\
        -\frac{1}{\sqrt{2}} [\delta b_x-i\delta b_y] & \frac{1}{\sqrt{2}}  [b_x-ib_y] & \text{J}+  b_z & 0 \\
        \frac{1}{\sqrt{2}} [\delta b_x+i\delta b_y] & \frac{1}{\sqrt{2}}  [b_x+i b_y] & 0 &  \text{J}-  b_z
    \end{pmatrix},
\end{align}
\end{widetext}

where the explicit dependence on $t$ has been removed to simplify the notation, $b_x = b_{x_1} + b_{x_2}$, $b_y = b_{y_1} + b_{y_2}$, $b_z = b_{z_1} + b_{z_2}$, $\delta b_x = (b_{x_1} - b_{x_2})$, $\delta b_y = (b_{y_1} - b_{y_2})$ and $\delta b_z = (b_{z_1} - b_{z_2})$. This Hamiltonian can be deduced from the Hamiltonian of the different interactions and from symmetry arguments (see Appendix~B). In this work, we focus on ST$_0$ qubits, where the computational basis are formed by the states $\ket{S} \equiv \ket{0}$ and $\ket{T_0}\equiv \ket{1}$, and the Hamiltonian is:

\begin{align} \label{Eq: ST0Hamil}
    \text{H}_{ST_0}=\begin{pmatrix}
        \text{J}&  \delta b_z \\
        \delta b_z &  -\text{J}
    \end{pmatrix}.
\end{align}

The rotation operator is defined as (See Appendix A):

\begin{align} \label{Eq: Rotation}
    R_{\hat{r}} (\theta_x + \theta_z)= \text{exp}\left(\frac{-i}{\hbar} \frac{\theta_x\sigma_x + \theta_z\sigma_z}{2}\right),
\end{align}

where the angles are $\theta_x =\tau_G \lambda_x= 2\tau_G \delta b_z$ and $\theta_z =\tau_G \lambda_z = 2\tau_G \text{J}$. The rotation axis is given by:

\begin{align*}
\hat{r}=\frac{\delta b_z \hat{x} + \text{J}\hat{z}}{\sqrt{J^2  +\delta b_z^2 }}.
\end{align*}

The readout of the ST$_0$ spin qubit is done by means of the Pauli spin blockade. We briefly explain how it works superficially for the sake of completeness, but we recommend the following works related to this topic~\cite{PhysRevB.76.035315,Overhauser,ReadoutDecay,ReadoutDecay2}. Due to the Pauli exclusion principle the electrons of the state $\ket{S}$ can tunnel into the (0,2) charge configuration; on the other hand, the state $\ket{T_0}$ is symmetric and the electrons can not occupy the same QD. This basic principle in quantum mechanics, allows us to differentiate the qubit state by tuning the potential barrier or by changing the detuning $\varepsilon$ between the QDs to allow the coherent transition. Finally, a charge measurement is performed to measure the state of the qubit.

\section{Time evolution and readout of ST$_0$ qubits } \label{Sec: Time Evolution}

\noindent The goal now is to determine the free and controlled time evolution of an ST$_0$ qubit when there are transverse magnetic fields, i.e. $\vec{b}_{x,y}(\vec{r},t) \neq 0$ and transitions between the computational basis and the $\ket{T_{\pm}}$ states are possible. In this scenario, the evolution of the system is described by the following Hamiltonian:
\begin{align}\label{Eq: Toy Model}
    \text{H}=\begin{pmatrix}
        \text{H}_{ST_0} & \text{H}_{leak} \\
        \text{H}_{leak}^\dagger & \text{H}_{T_{\pm}}
    \end{pmatrix}.
\end{align}

\noindent The computational basis is given by the Hamiltonian $\text{H}_{ST_0}$ defined in Eq.9 for $\delta b_z = 0$ and the leakage Hamiltonian $\text{H}_\text{leak}$ is defined as:
\begin{align}\label{Eq: Hleakage}
    \text{H}_\text{leak}=\begin{pmatrix}
          -\frac{1}{\sqrt{2}}  [\delta b_x+i\delta b_y] & \frac{1}{\sqrt{2}}  [\delta b_x-i\delta b_y] \\
           \frac{1}{\sqrt{2}}  [b_x+ib_y] & \frac{1}{\sqrt{2}}  [b_x- i  b_y] 
    \end{pmatrix}.
\end{align}

The $\text{H}_{T_{\pm}}$ Hamiltonian is the subspace of the states $\ket{T_\pm}$ and $\text{H}_{leak}$ by the transition's amplitudes from the computational basis to the $T_\pm$ states, see Eq.\ref{Eq: Hamil6x6}. In absence of leakage terms, i.e. $\text{H}_{leak}=0$, the computational basis ($\ket{S}$ and $\ket{T_0}$) are eigenstates, but in presence of transverse magnetic fields it is not the case. The eigenvalue of each eigenstate determines its evolution due to the Schr\"odinger equation. In presence of leakage, qubit rotations belongs to a $4$-dimensional Hilbert space and the expression of a rotation using the generators of the space is more complex (See Appendix~\ref{SecA: symmetry}). We use perturbation theory and a Dyson series expansion to study the dynamics of the system, i.e. to compare how the dynamics of the system change in presence of weak fields. In the next section, we analyze and compare the dynamics of the ST$_0$ qubit in different scenarios. 

\subsection{Time Evolution Simulations}

\noindent In this section, we present a numerical analysis of the dynamics of the system. We refer to the population of a certain state $ \ket{\phi}$ while evolving as $\text{Pop}\left(\phi,t\right)$. Assuming $\ket{\psi(0)}$ to be the initial state, the evolution of the previously mentioned population is given by:
\begin{align} \label{Eq: Population}
    \text{Pop}\left(\phi,t\right)&= |\langle \phi \ket{\psi(t)}|^2.
\end{align}

\noindent The dynamics is governed by the Hamiltonian given in Eq.~\ref{Eq: Hamil6x6}. There are different physical devices where a QD can be generated: SiMOS structures, GaAS and Si/SiGe are commonly used~\cite{RevModPhysSiQ}. The gyromagnetic ratio g , the effective mass of the electrons $m_e^{eff}$ and the relative permittivity $\epsilon_r$ depends on the semiconductor used to generate the QD. In addition, there are different ways to generate the electric and magnetic fields to generate the confinement and the Zeeman splitting, and different techniques to control the electron's spin state~\cite{QuantumDots}: Electron Spin Resonance (ESR), through the interaction of the electron with an external magnetic field; or Electric Dipole Spin Resonance (EDSR), through the interaction of the electron with an inhomogeneous magnetic field or the spin-orbit interaction. Furthermore, the values of the exchange coupling between the electrons depends on the potential detuning $\varepsilon$ and the effective tunneling amplitude $T_c$, which also depends on the shape of the QDs and the material. Generally, they are related by: $J \simeq 8\frac{4 T_c^2}{\sqrt{\varepsilon^2 + 4T_c^2}+\varepsilon  }$ but the precise values of $\varepsilon$ and $T_c$ are not the same across different technologies. Therefore, since the main objective of this work is to see which is the effect of leakage in the unitary evolution of a ST$_0$ qubit from a theoretical point of view, we set the values of $J$ and $ \delta b_z$ to 1 and $ b_z = 5$; while the values of the transverse magnetic fields are given as a percentage of the $J$ value. Also, we fix the value of $\hbar$ to 1.

\begin{figure}[t!]
    \centering
    \subfloat[]{\includegraphics[clip,width=.95\columnwidth]{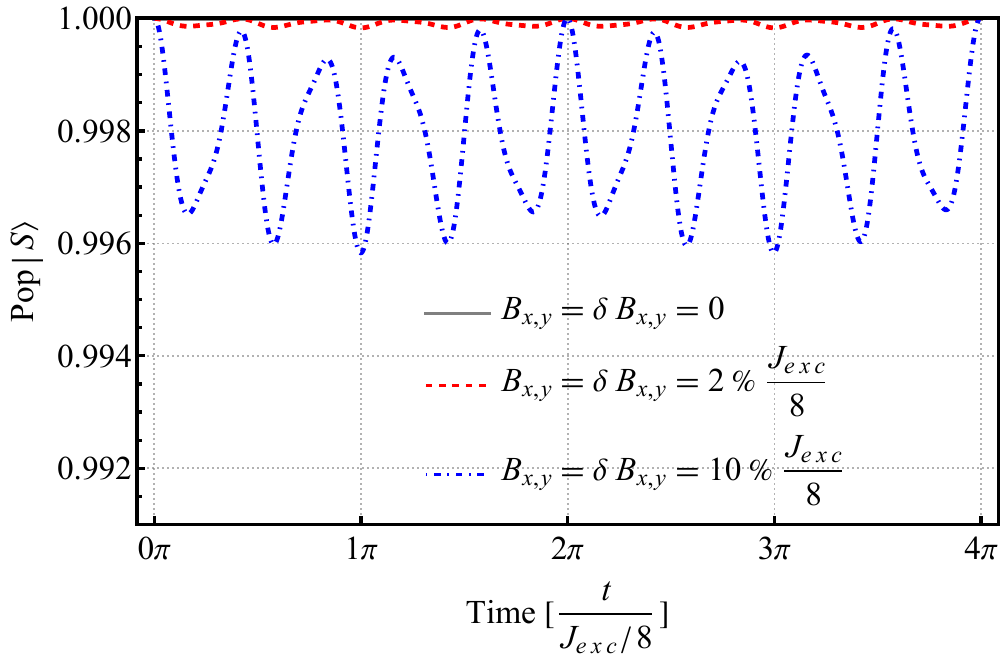}}
    
    \subfloat[]{\includegraphics[clip,width=.95\columnwidth]{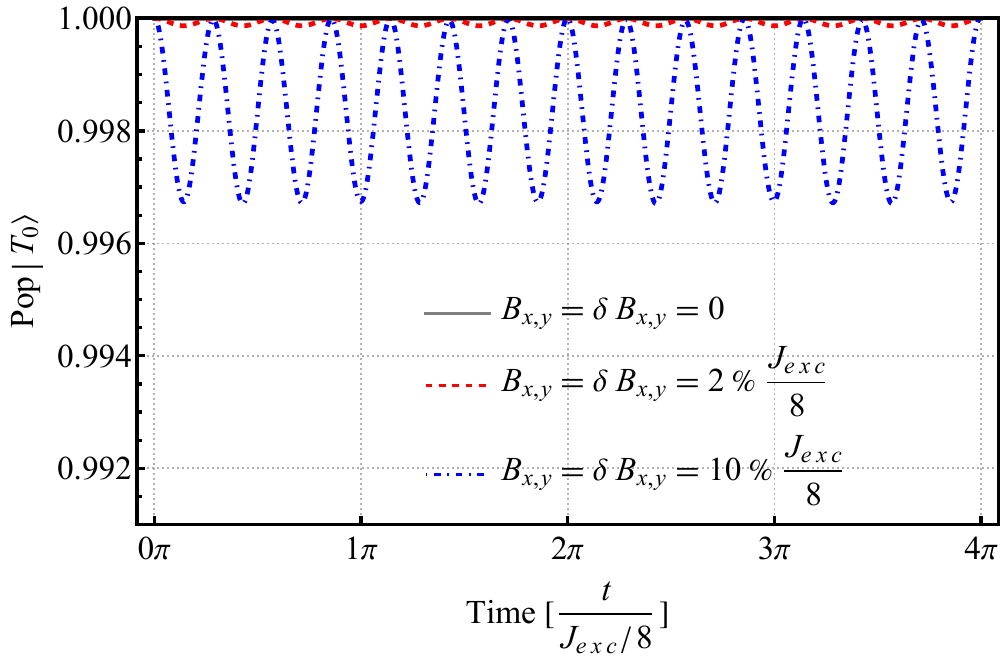}}
    
    \caption{Free evolution of (a) $\ket{S}$ and (b) $\ket{T_0}$ states in presence of leakage.The values of the transverse magnetic fields have been taken relative to the normalized value of $J = 1$, such that the red (dotted) line refers to transverse fields at 2\% and the blue (dash-dotted) line to 10\%. We can observe that the population remains constant when there is not magnetic transverse fields since the states are eigenstates of the Hamiltonian (Eq.~\ref{Eq: Toy Model}), however in presence of them the population varies over time.}
    \label{fig: Free evolution}
\end{figure}

\subsubsection{Free Evolution} \label{Sec: Free evolution}

\noindent The free evolution of the system is governed by the diagonal Hamiltonian Eq.~\ref{Eq: ST0Hamil} for $\delta b_z = 0$. Without a transverse magnetic field, i.e. $\vec{b}= \left(0,0,b_z \right)$, the singlet and triplet states are eigenstates of the Hamiltonian and their populations remain constant. Therefore, the Hamiltonian of the tetradimensional Hilbert space is defined as

\begin{align} \label{Eq: HEnergy}
    \text{H}_\text{Energy}&= \text{J}\eta+  \left( \frac{b_z}{2}\Lambda_3 + \frac{\sqrt{3}b_z}{2} \Lambda_8 \right) \nonumber \\
    & \hspace{-0.5cm}=  \small\begin{pmatrix}
         -\text{J} & 0 & 0 & 0 \\
         0 & \text{J}& 0 & 0 \\
         0 & 0 & \text{J}+ b_z & 0 \\
         0 & 0 & 0 & \text{J}-b_z     
    \end{pmatrix},
\end{align}
where $\eta = \text{diag}(-1,+1,+1,+1)$,  $\Lambda_3$ and $\Lambda_8$ are Gell-Man matrices (See Appendix~\ref{SecA: symmetry}); and we reorganized the terms to keep the order used in Eq.~\ref{Eq: Toy Model}. The 4 eigenvalues are  $\{\lambda_{\ket{S}}= -\text{J}$, $\lambda_{\ket{T_0}}=J$, $\lambda_{\ket{T_+}}=J+ b_z$, $\lambda_{\ket{T_-}}=J- b_z\}$. 

On the other hand, the singlet and triplet states are not eigenstates of the system when there is an external transverse field, so their populations vary over time. The new eigenstates $\{\varphi_1,\varphi_2,\varphi_3,\varphi_4\}$ can be obtained by diagonalizing the new Hamiltonian and the singlet and triplet states can be represented as a linear combination of them:
\begin{subequations}\label{Eq: Comp eigenstates}
    \begin{align} 
    \ket{S}&= \sum_i \alpha_i \ket{\varphi_i}, \label{Eq: Comp eigenstates1} \\
    \ket{T_0}&= \sum_i \beta_i \ket{\varphi_i}, \label{Eq: Comp eigenstates2} \\
    \ket{T_+}&= \sum_i \nu_i \ket{\varphi_i},  \label{Eq: Comp eigenstates3}\\
    \ket{T_-}&= \sum_i \kappa_i \ket{\varphi_i}. \label{Eq: Comp eigenstates4}
\end{align}
\end{subequations}

\noindent Each eigenstate has an associated eigenvalue $\{\lambda_1,\lambda_2,\lambda_3,\lambda_4\}$ and their time evolution operators are defined by Each eigenstate has an associated eigenvalue $\{\lambda_1,\lambda_2,\lambda_3,\lambda_4\}$ and their time evolution operators are defined by the solution of the Schrödinger equation.

\noindent Then, when we prepare an arbitrary state  $\ket{\psi}=\xi_1 \ket{\varphi_1}+ \xi_2  \ket{\varphi_2}+ \xi_3  \ket{\varphi_3}+ \xi_4  \ket{\varphi_4} $ and let it evolve over time, the temporal evolution of the population of certain state $\ket{\phi}=a_1 \ket{\varphi_1}+ a_2 \ket{\varphi_2}+ a_3 \ket{\varphi_3}+ a_4 \ket{\varphi_4}$ is given by:
\begin{align}
    \text{Pop}(\phi,t)=& |\langle \phi \ket{\psi(t)}|^2 \nonumber \\ 
    =&|a_1^* \xi_1 e^{-i\lambda_1 t} + a_2^* \xi_2 e^{-i\lambda_2 t} \nonumber \\
    &+ a_3^* \xi_3 e^{-i\lambda_3 t}+a_4^* \xi_4 e^{-i\lambda_4 t}|^2.
\end{align}

The external transverse fields generate a mixing between the computational basis ($\ket{S}$, $\ket{T_0}$) and the other triplets ($\ket{T_\pm}$). It causes oscillations between the computational subspace and the leakage subspace. Since a fraction of the computational states' population is transferred to $\ket{T_\pm}$, the unitary evolution of their population oscillate with an associated frequency of:

\begin{align}
\Omega_{S-T_\pm}&= 2\sqrt{ \left(\frac{\lambda_{S} - \lambda_{T_\pm}}{2}\right)^2+ \left|\frac{-1}{\sqrt{2}}  [\delta b_x \pm i \delta b_y]\right|^2},\\
\Omega_{T_0-T_\pm} &= 2\sqrt{ \left(\frac{\lambda_{T_0} - \lambda_{T_\pm}}{2}\right)^2+ \left|\frac{1}{\sqrt{2}}  [b_x\pm ib_y]\right|^2}.
\end{align}

The energy gap between the state $\ket{T_0}$ and the states $\ket{T_\pm}$ is the same, so they oscillate coherently, as can be observed in Fig.~\ref{fig: Free evolution} panel b. However, the energy gap between the state $\ket{S}$ and the states $\ket{T_\pm}$ is not symmetric causing a dephasing between both transitions, see Fig.\ref{fig: Free evolution} panel a.
   
\subsubsection{Rotations}

\noindent In the following, we analyze the dynamics of the ST$_0$ qubit when we apply external magnetic fields to induce single-qubit gates and, thus, study the impact of leakage on such rotations.

To perform rotations over different axes, we need to apply magnetic fields with different amplitudes and directions. Global magnetic fields do not break the SU(2) $\otimes$ SU(2) symmetry, By global we refer to the scenario in which the magnetic field is equal in both QDs. However, the components in the $\hat{x}$ and/or $\hat{y}$ of the external magnetic field break the spin angular momentum symmetry in direction $\hat{z}$ ($S_z$) and allow transitions between levels with a difference in angular momentum $\pm1$ inside the SU(3) subspace (recall that SU(2) $\otimes$ SU(2) = U(1) $\oplus$ SU(3)). Also, the SU(2) $\otimes$ SU(2) symmetry is broken when there exists a non-trivial gradient on the magnitudes of the external transverse magnetic fields, allowing transitions between the U(1) and SU(2) subspaces.

\begin{figure}[t!]
    \includegraphics[width=0.9\columnwidth]{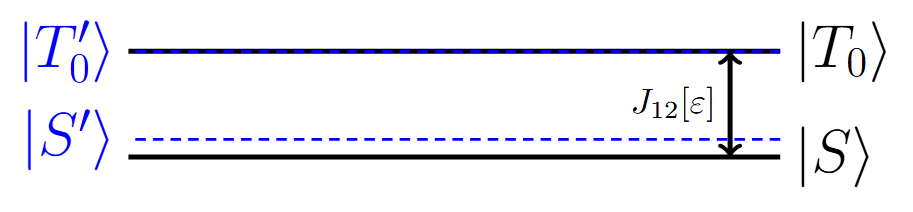}
    \caption{Schematic representation of the computational states' energy shift due to transitions to higher energy levels using perturbation theory.}
    \label{fig: ST0 Pert}
\end{figure}

In this section, we continue assuming a constant value of the fields during the gate time. The Hamiltonian defining the singlet-triplet states (Eq.~\ref{Eq: HEnergy}) and the time dependent rotation Hamiltonian depend on the values of the QD's potentials and the external magnetic fields. The dynamics is then defined by the following Hamiltonian:

\begin{align}
    \text{H}_\text{Gate} = \text{H}_\text{Energy} + \text{H}_R(t),  
\end{align}
where $\text{H}_R(t)$ is defined by the Hamiltonian given in Eq.~\ref{Eq: Hamil6x6}. First, we study rotations around the $\hat{z}$ axis induced by applying an external electric field that changes the exchange coupling, $J$. Such rotations change the relative phase between the states $\ket{S}$ and $\ket{T_0}$, but do not induce transitions between them. Given a general state $\ket{\psi} = \alpha \ket{S} + \beta \ket{T_0}$, the relative phase change over time is given by:
\begin{align} \label{Eq: dephasing}
    \ket{\psi(t)} &= \alpha \hat{U}_{\lambda_1}(0,t)\ket{S} + \beta \hat{U}_{\lambda_2}(0,t) \ket{T_0} \nonumber\\
    &= \hat{U}_{\lambda_1}(0,t) \left(  \alpha \ket{S} + \beta \frac{\hat{U}_{\lambda_2}(0,t)}{\hat{U}_{\lambda_1}(0,t)} \ket{T_0}\right) \nonumber\\
    &=\alpha \ket{S} + \beta e^{\frac{-i}{\hbar} \left(\lambda_{\ket{T_0}} - \lambda_{\ket{s}}\right) t} \ket{T_0},
\end{align}
where $\lambda_{\ket{S}}=  -\text{J}$ and $\lambda_{\ket{T_0}} = \text{J}$ are the eigenvalues of the Hamiltonian. Note that the value of $J$ changes during the gate time $\tau_G$ due to the variation of $\varepsilon$. As we stated before, when there is an external transverse magnetic field that may induce leakage, we can express the singlet-triplet states as a linear combination of the new eigenstates of the Hamiltonian $\{\phi_1,\phi_2,\phi_3,\phi_4\}$, as in Eq.~\ref{Eq: Comp eigenstates}, with $\alpha'_i, \beta'_i, \nu'_i, \kappa'_i$ the complex amplitudes and $\{\lambda'_1,\lambda'_2,\lambda'_3,\lambda'_4\}$ the new eigenvalues.

To mathematically analyze how leakage changes the dynamics, we use perturbation theory (PT) since we are working in the weak transverse field regime, i.e. $\bigl|\pm \frac{1}{\sqrt{2}} (\delta b_x \pm i \delta b_y)\bigr|$, $\bigl|\pm \frac{1}{\sqrt{2}} ( b_x \pm i b_y)\bigr| << \bigl|J\bigr|$. We can split the Hamiltonian in two different parts: The unperturbed Hamiltonian $\text{H}_0$ given in Eq.~\ref{Eq: HEnergy}, and the perturbed Hamiltonian $\text{H}_p$ formed by the non-diagonal terms. Then, the new eigenvalues can be approximated as:

\begin{subequations} \label{Eq: PTLambdas}
    \begin{align} 
    \lambda'_1  &\approx \lambda_1+ \sum_m \frac{\bra{S}\text{H}_I \ket{m} \bra{m}\text{H}_I \ket{S}}{\lambda_1 - \lambda_m},  \\
    \lambda'_2  &\approx \lambda_{2} + \sum_m \frac{\bra{T_0}\text{H}_I \ket{m} \bra{m}\text{H}_I \ket{T_0}}{\lambda_{2 }- \lambda_m} ,\\
    \lambda'_3  &\approx \lambda_{3} + \sum_m \frac{\bra{T_+}\text{H}_I \ket{m} \bra{m}\text{H}_I \ket{T_+}}{\lambda_{3} - \lambda_m}, \\
    \lambda'_4  &\approx \lambda_{4}+ \sum_m \frac{\bra{T_-}\text{H}_I \ket{m} \bra{m}\text{H}_I \ket{T_-}}{\lambda_{4} - \lambda_m}.
    \end{align}
\end{subequations}

\noindent In theory, the perturbation could make the energy gap between $\lambda_1$ and $\lambda_2$ larger or smaller, but for the DQD system it only reduces the value of $\lambda_1$. The shift of the eigenvalues of the computational states is represented graphically in Fig.~\ref{fig: ST0 Pert}.

Now, given an arbitrary state $\ket{\psi}= \alpha \ket{\phi_1} + \beta \ket{\phi_2}$, it evolve as (see Eq.~\ref{Eq: dephasing}):

\begin{align}
    \ket{\psi(t)}= \alpha \ket{\phi_1} + \beta e^{\frac{-i}{\hbar} \left(\lambda'_2 - \lambda'_1\right) t} \ket{\phi_2}.
\end{align}

Then, the time to perform the desired relative phase change for the computational basis depends not only on the values of $J$, but also on the values of the transverse magnetic field. If $\left(\lambda'_2 - \lambda'_1\right) > \left(\lambda_{\ket{T_0}} - \lambda_{\ket{s}}\right)$ for a given external magnetic field $\vec{B}$, the time required to perform a rotation would decrease, and viceversa. However, in the case of a small perturbation, the eigenvalue $\lambda'_2$ remain unchanged while eigenvalue $\lambda'_1$ is always larger than $\lambda_{\ket{s}}$, thus, $\left(\lambda'_2 - \lambda'_1\right) \leq \left(\lambda_{\ket{T_0}} - \lambda_{\ket{s}}\right)$ for any amplitude of the external transverse magnetic fields.

To perform rotations around the $\hat{x}$ axis, we have to vary the external magnetic field in the $\hat{z}$ direction to create a non-trivial magnetic field gradient between the QDs, i.e. $\delta b_z \neq 0$. Now, the situation is completely different because this field is not in the weak regime and we cannot use PT to study the effect of leakage on the dynamics. In the case of ST$_0$ qubits without leakage, a rotation was defined in Eq.~\ref{Eq: Rotation}; which in the presence of leakage is given by:
\begin{widetext}
    \begin{align} \label{Eq: RotationDim4}
        R(\text{H},\tau_g) &= \mathcal{T} \biggl\{ \exp\biggl( \frac{-i}{2} \int_0^{\tau_G} dt \biggl[J \eta +   \frac{b_z}{2}\Lambda_3 + \frac{\sqrt{3}b_z}{2} \Lambda_8 + \frac{1}{\sqrt{2}} b_x \Lambda_1 + \frac{1}{\sqrt{2}} b_x \Lambda_6 + \frac{1}{\sqrt{2}} b_y \Lambda_2 + \frac{1}{\sqrt{2}} b_y \Lambda_7  \nonumber\\
         &\hspace{4cm}+   \delta b_z \Lambda'_3 - \frac{1}{\sqrt{2}} \delta b_x \Lambda'_1 + \frac{1}{\sqrt{2}} \delta b_x \Lambda'_3 + \frac{1}{\sqrt{2}} \delta b_y \Lambda'_2+\frac{1}{\sqrt{2}} \delta b_y \Lambda'_6) \biggr]\biggr) \biggr\},
    \end{align}
\end{widetext}
where $\Lambda_i$ are the Gell-mann matrices and $\Lambda_i'$ are rotations that break the SU(2) $\otimes$ SU(2) symmetry. This equation is explained and deduced in Appendix~\ref{SecA: symmetry}.

We study the rotations induced by the Hamiltonian using the time ordering operator and expand it using a Dyson series (see Appendix~\ref{SecA: Dyson series}). We get the evolution operators for transitions $\ket{S}\xrightarrow{} \ket{T_0}$ and $\ket{T_0}\xrightarrow{} \ket{S}$, which are given by the same expression since $\bra{S}\text{H}_I(t)\ket{T_0}= \left(\bra{T_0}\text{H}_I^\dagger(t)\ket{S}\right)^\dagger = \left(\bra{T_0}\text{H}_I(t)\ket{S}\right)^\dagger$, using the hermitian property of the Hamiltonian $\text{H}=\text{H}^\dagger$. To work with the Dyson series we split the Hamiltonian in its non-interacting or diagonal terms given in Eq.~\ref{Eq: HEnergy} and in its interacting or off-diagonal terms: 
\begin{align}
    \text{H}_I(t) =  \begin{pmatrix}
        0 & \\
         & \text{H}_{I\{T_0,T_\pm\}}
    \end{pmatrix} + \text{H}_{I\{BS\}},
\end{align}
where:
\begin{align}
    \text{H}_{I\{T_0,T_\pm\}} =  {\scriptsize\begin{pmatrix}
        0 & \frac{1}{\sqrt{2}} \left(b_x -i b_y\right) & 0 \\
        \frac{1}{\sqrt{2}} \left(b_x +i b_y\right) & 0 & \frac{1}{\sqrt{2}} \left(b_x -i b_y\right) \\
        0 & \frac{1}{\sqrt{2}} \left(b_x +i b_y\right) & 0
    \end{pmatrix},}
\end{align}

and

\begin{align} \label{Eq: HintBS}
    \text{H}_{I\{BS\}} =& {\scriptsize\begin{pmatrix}
         0 &  \frac{-1}{\sqrt{2}} (\delta b_x + i\delta b_y) & \delta b_z   & \frac{1}{\sqrt{2}}(\delta b_x - i\delta b_y) \\
         \frac{-1}{\sqrt{2}}(\delta b_x - i\delta b_y) & 0 & 0 & 0 \\
         \delta b_z & 0 & 0 & 0 \\
          \frac{1}{\sqrt{2}}(\delta b_x + i\delta b_y) & 0 & 0 & 0  
    \end{pmatrix}.}
\end{align} 

\noindent Both expressions are deduced in Appendix~\ref{SecA: symmetry}. The unitary evolution operator when only the term $\delta B_z \neq 0$ is given by:
\begin{align} \label{Eq: RotDyson}
    \hat{U}_I(\tau_G,0)&=\exp\biggl(\frac{-i\tau_G}{\hbar}  \delta b_z \sigma_x \biggr).
\end{align}

\noindent However, whenever we apply a general magnetic field, using Eq.~\ref{Eq: RotationDim4} and a Dyson series, we can rewrite approximately the unitary evolution of a rotation around $\hat{x}$ as:
    \begin{align}\label{Eq: RotDysonleak}
        \hat{U}_I(t,0)\approx &\frac{-it}{\hbar} \ \delta b_z
        -i\left(\frac{t}{\hbar}\right)^2 \frac{1}{2}  (\delta b_x b_y-\delta B_y B_x) 
\end{align}

\noindent The phase change between the time evolution of the system with and without leakage is given by the difference in the argument of the unitary operators in Eq.~\ref{Eq: RotDyson} and in Eq.~\ref{Eq: RotDysonleak}. Furthermore, we can compute the effective amplitudes of the transitions at second order approximation as:
\begin{align}
    A_{\ket{S} \rightarrow{} \ket{T_0}} =&  \delta b_z \nonumber\\
    & + \frac{   \left(b_x -i b_y \right) \left( \delta b_x + i \delta b_y \right)}{E_{\ket{S}} - E_{\ket{T_-}}}\nonumber\\
    & +\frac{   \left(b_x +i b_y \right) \left( \delta b_x - i \delta b_y \right)}{E_{\ket{S}} - E_{\ket{T_+}}}
\end{align}
and
\begin{align}
    A_{\ket{T_0} \rightarrow{} \ket{S}} =&  \delta b_z \nonumber\\
    &+ \frac{   \left(b_x +i b_y \right) \left( \delta b_x - i \delta b_y \right)}{E_{\ket{T_0}} - E_{\ket{T_-}}} \nonumber\\
    & +\frac{   \left(b_x -i b_y \right) \left( \delta b_x + i \delta b_y \right)}{E_{\ket{T_0}} - E_{\ket{T_+}}}.
\end{align}

\begin{figure}[t!]
    \centering
    \subfloat[]{\includegraphics[width=1.02\columnwidth]{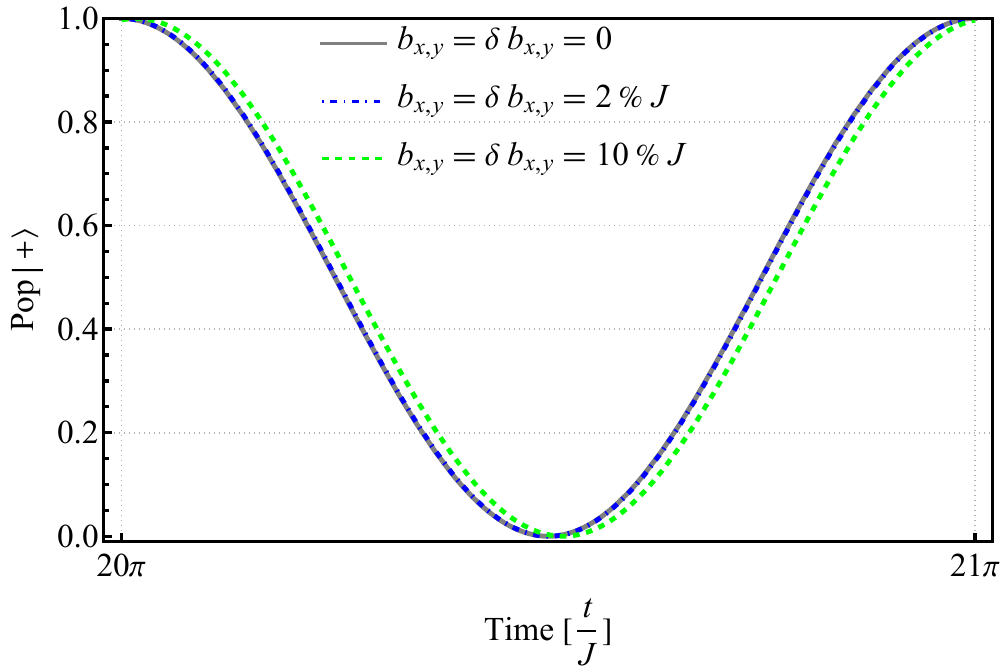}}
    
    \subfloat[]{\includegraphics[width=1.\columnwidth]{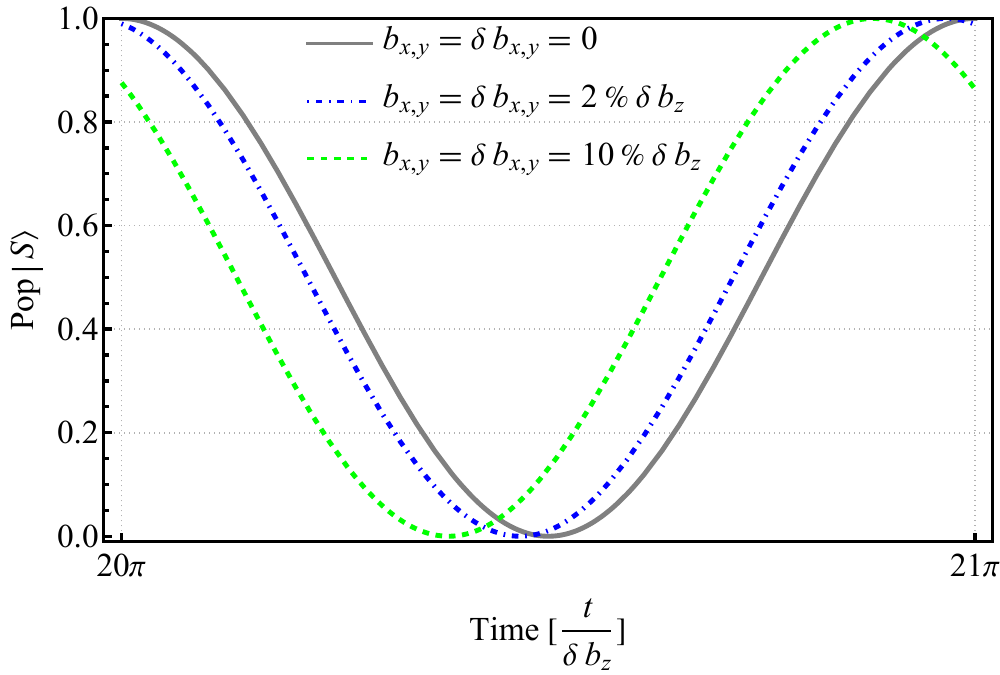}}
    \caption{Comparison of the time evolution of the states (a) $\ket{+}$ when a rotation around $\hat{z}$ is performed for $\text{J}=1$ and (b) $\ket{S}$ when we apply a magnetic field such as $\delta b_z = 1$. The gray (solid) line is the evolution without leakage terms described by Eq.10, the blue (dashed) and the green (dashed-dotted) lines are the evolution for the values of the transverse magnetic field $b_{x,y}=\delta b_{x,y} = 0.02$ and $b_{x,y}=\delta b_{x,y} = 0.1$, respectively.}
    \label{fig: Heff}
\end{figure}

Now, we can write the effective Hamiltonian of the dynamics of the ST$_0$ qubit as:

\begin{align} \label{Eq: Heffective}
    H_{\text{eff}}(t) = \begin{pmatrix}
         \lambda'_1&  A_{\ket{T_0} \rightarrow{} \ket{S}}  \\
         A_{\ket{S} \rightarrow{} \ket{T_0}} & \lambda'_2
    \end{pmatrix},
\end{align}
where $\lambda'_1$ and $\lambda'_2$ are given in Eq.~\ref{Eq: PTLambdas}. Finally, we can compute the rotation operators as $R(H_{\text{eff}}(t), \tau_g)$ using Eq.~\ref{Eq: Rotation} instead of resolving the rotation in the Hilbert space of dimension 4 given in Eq.~\ref{Eq: RotationDim4}. In Fig.~\ref{fig: Heff}, we compare the evolution of the system without leakage with the evolution defined by the effective Hamiltonian in presence of a transverse magnetic fields using the Eq.~\ref{Eq: Rotation}. The evolution of the state $\ket{+}=\frac{\ket{S}+\ket{T_0}}{\sqrt{2}}$ during the rotation is shown in the panel a for different values of the transverse magnetic fields. We observe that the rotation in presence of leakage terms is slower than the rotation without them , i.e. there is a phase shift between both evolutions due to the change of the energies. This phase change grows when the amplitudes of the transverse magnetic fields increase or when the energy gap between the states $\ket{T_{\pm}}$ and the computational states is reduced. Similarly, in the panel b the evolution of $\ket{S}$ is shown for different values of the transverse magnetic field, $\delta b_z =1$ and $\text{J}=0$. Furthermore, it gives us the freedom to control the speed of the time evolution operators by means of the signs and amplitudes of the transverse magnetic fields. Comparing the effective time evolution operators with the one given by the Hamiltonian in Eq.~\ref{Eq: ST0Hamil}, it becomes evident how these terms induce a change in the system's dynamics. Explicitly, the required time to perform an X and a Z rotation has a dependence on the transverse magnetic field given by: 

\begin{align}
    \tau_{X} & \simeq  \frac{\pi}{\sqrt{\delta b_z^2 +  4\frac{ b_y^2 \delta b_x^2 + b_x^2 \delta b_y^2}{bz^2}}}, \label{Eq: timeX} \\
    \tau_Z & \simeq \frac{\pi}{ \text{J}_{exc}/4- \left(\delta b_x^2 + \delta b_y^2\right) \frac{\frac12  \text{J}_{exc}}{b_z^2 -\frac{ \text{J}_{exc}^2}{16}}},\label{Eq: timeZ}
\end{align}

for $J=0$ and for $\delta b_z = 0$ for rotations around $\hat{x}$ and $\hat{z}$, respectively. This difference accumulates throughout the system's evolution, producing a coherent error during computation that depends on the system's energy landscape. Although small, it is important to consider this effect in algorithms requiring the implementation of a large number of gates, such as Shor's algorithm~\cite{Shor}. Furthermore, the significance of the energy gap between the computational basis and states outside of it is demonstrated. It is paramount to note that in the absence of a longitudinal gradient $\delta b_z$ remains closed and is formed by the new eigenstates of the Hamiltonian. However, once a finite $\delta b_z$ is present the transverse field leaves residual couplings to $\ket{T_\pm}$. These couplings slightly shift the logical energies and manifest themselves as the coherent phase shifts discussed in this section. 

\subsubsection{Fidelity} \label{Sec: fidelity}

In spin qubits in QDs devices the primary sources of errors are the charge noise which generates an small uncertainty to the electric fields; and hyperfine interactions between the spin of the electrons and the spin of the surrounding atoms~\cite{Charge1, NoiseMicro}. In this section we analyze the impact on the gate fidelity in presence of noise and a transverse magnetic field. To take the noise into account we redefine the fields as: 
\begin{align}
    &\text{J} \xrightarrow{}\text{J} + \widetilde{\text{J}} \\
    &b_z \xrightarrow{} b_z + \widetilde{b_z} \\
    &\delta b_z \xrightarrow{} \delta b_z + \widetilde{\delta b_z}
\end{align}

where the fields denoted by $ \widetilde{ \cdot}$ follow a probability distribution. Usually, in QDs the charge noise has a power spectral density of $1/f$ and the Hyperfine induced Overhauser field fluctuations are treated as gaussian noise. We neglect the effect of such gradients to the transverse fields since we are already defining them as weak fields. Therefore, we can redefine the ST$_0$ Hamiltonian as:

\begin{align}
    \text{H}_\text{noisy} = \text{H} + \text{H}_n
\end{align}, 

where H is given in Eq.~\ref{Eq: Hamil6x6} and $\text{H}_n$ is given by:
\begin{align}
    \text{H}_n &= \begin{pmatrix}
        -\frac{\widetilde{  J_{exc}}}{8} & \widetilde{\delta b_z} & 0 & 0 \\
         \widetilde{\delta b_z} & \frac{\widetilde{ J_{exc}}}{8} & 0 & 0 \\
         0 & 0 & \frac{\widetilde{ J_{exc}}}{8} +\widetilde{ b_z}& 0 \\
         0 & 0 & 0 & \frac{\widetilde{ J_{exc}}}{8}-\widetilde{b_z}  
    \end{pmatrix}.
\end{align}

The average gate fidelity of a noisy gate $R_G$ generated by the Pauli generator G, is defined as~\cite{avgfid}:
\begin{align} \label{Eq: Fidelity}
    \langle \mathcal{F}_{R_G} \rangle = \int d \psi |\bra{\psi} \hat{U}_{Id}^\dagger \hat{U}_\text{noisy}(t)\ket{\psi}|^2,
\end{align}
where $\hat{U}_{Id}$ is the ideal unitary evolution of the gate given by $\textrm{H}$; and $\hat{U}_\text{noisy}$ is the noisy evolution given by $\textrm{H} + \textrm{H}_n $. It is top bounded by the expression~\cite{van2024fidelity}:
\begin{align} \label{Eq: FidBound}
    \langle \mathcal{F}_{R_G} \rangle\geq \frac{|Tr(\hat{U}_{n}\hat{U}_{Id}^\dagger)|^2 + d }{d(d+1)},
\end{align}
where $d$ is the dimension of the Hamiltonian. After some algebra and under the assumption that the uncertainty of the fields due to the noise is small compared with their amplitudes, the gate fidelities with and without leakage terms are approximately given by:

\begin{align}
    \langle \mathcal{F}_{R_G} \rangle= \langle \mathcal{F}^{leak}_{R_G} \rangle&\geq  1 - \frac{\left( 3\frac{\widetilde{  J_{exc}}^2}{16} + 8 \widetilde{  \delta b_z}^2\right)  \tau_G^2 }{20},
\end{align}

where $\tau_G$ is the gate time given in Eq.~\ref{Eq: timeX} and Eq.~\ref{Eq: timeZ}. Therefore, the fidelity is not directly affected by weak transverse fields. Charge noise and fluctuations of the magnetic field induce decoherence within the computational basis, while leakage into auxiliary states is significantly suppressed. This suppression arises because the energy gap between the computational states and the leakage levels is substantially larger than the noise amplitude. However, the presence of leakage induce new paths between the computational basis and modify the gate time $\tau_G$, inducing coherent errors which accumulates along the quantum computation. As a result, while the fidelity bound remains approximately the same, the total gate error depends on the transverse field through its effect on the gate duration. The given bound is also valid for two-qubit gates for LD spin qubits since the entangling gates are realized by means of the Hamiltonian given in Eq.~\ref{Eq: 2qubit}, hence they are subject to the same underlying physics~\cite{ST0Control}.

\subsection{Readout process} \label{Sec: Readout}

In this section we analyze the impact of a transverse magnetic field during the readout process. The readout process in its most general case is composed of two steps, where the first one is a coherent evolution from the single electron occupation to the double occupation configuration. Only the singlet state population can be transferred to this configuration due to Pauli exclusion principle. Finally, a charge measurement is performed to readout the state of the ST$_0$ qubit. A source of readout errors is the fluctuations of the Overhauser field gradient, which arises because the environment of each QD creates a longitudinal magnetic field gradient $\Delta b_z$. It allows the indirect decay $\ket{T_0(1,1)}\xrightarrow[]{}\ket{S(0,2)}$ generating a mixing between the computational basis. It is important to note that it can be suppressed by increasing the detuning $\varepsilon$~\cite{Overhauser}. Here, we analyze the effect of a transverse global magnetic field $b_{x,y}$ on the fidelity of the readout process, based on the longer lifetime of the states $\ket{T_\pm}$~\cite{ReadoutDecay}. The same effect has been studied in~\cite{StrongRegime} when the amplitude of the transverse magnetic field is comparable to the energy splitting between the levels.

Along this section, we will work in the interaction picture and we will define the following Hamiltonians:

\begin{align}
    \textrm{H}_{ro}^I &= t_c \left( \ket{S(1,1)} \bra{S(0,2)} + h.c. \right), \\
    \textrm{H}_{ro-leak}^I &= t_c \left( \ket{S(1,1)} \bra{S(0,2)} + h.c. \right) \\
    &+ \left[ \frac{1}{\sqrt{2}}(b_x + i b_y) \ket{T_0}\bra{T_+} + h.c. \right] \\
    &+ \left[ \frac{1}{\sqrt{2}}(b_x - i b_y) \ket{T_0}\bra{T_-} + h.c. \right] \\
    \textrm{H}_n^I &= \widetilde{\delta b_z} \left( \ket{S(1,1)} \bra{T_0} + h.c. \right),
\end{align}

where $t_c$ is the tunneling coupling, $\textrm{H}_{ro}^I$ and $\textrm{H}_{ro-leak}^I$ are the interaction Hamiltonians which allow the transition from the (1,1) to the (0,2) configuration without and with leakage, respectively; and $\textrm{H}_n^I$ is the Hamiltonian of the noise which causes a mixing between the computational states. There is a possible avoided crossing point between the singlet and the $\ket{T_+}$ state ( or $\ket{T_-}$; depending on the device's material) during the readout process. The mixing between these states is indeed mediated by the gradient of the transverse field $\delta b_{x,y}$. Even though small transverse magnetic field gradients may exits, the  corresponding transition frequency is given by:

\begin{align}
    \Omega_{S(0,2)-T_-} = 2\sqrt{\Delta(\varepsilon)- \delta b_\perp^2},
\end{align}

where $\Delta(\varepsilon)$ is the energy difference between the singlet and the triplet. The minimum energy gap at the avoided crossing is given by $2|\delta b_\perp^2|$. However, in the weak magnetic field regime, the coherent oscillation period is $\tau= \Omega_{S(0,2)-T_-}^{-1}$ caused by small oscillations of the magnetic field amplitudes which is much longer than the relaxation time $T_1$ governing the decay~\cite{T2noise}. Consequently, the coherent coupling is too weak to drive noticeable coherent oscillations during the readout process. The average gate fidelity defined in Eq.~\ref{Eq: Fidelity} and bounded by Eq.~\ref{Eq: FidBound}  of the readout process, with and without leakage, is given by:

\begin{align}
    \langle \mathcal{F}_{readout} \rangle &\geq \frac{\left[ 5 -  \widetilde{\delta b_z}^2\tau^2 \right]^2+ 5}{30} \simeq 1 - \frac13 \widetilde{\delta b_z}^2 \tau^2. \\
    \langle \mathcal{F}_{readout}^{leak.} \rangle &\geq  \frac{\left[5-\widetilde{\delta b_z^{eff}}^2 \tau^2\right]^2+5}{30} \simeq 1 - \frac13 \widetilde{\delta b_z}^2 \alpha^2\tau^2,
\end{align}

with $\widetilde{\delta b_z^{eff}}^2= \alpha^2 \widetilde{\delta b_z}^2$, $\alpha = \left( 1+ \frac{B_x^2 + B_y^2}{b_z^2} + \frac{B_x^2 + B_y^2}{b_z^2}\right)^{-1/2}$ and $\tau$ is the time. We can relate this process with a Gaussian noise model, where the coherence time of the state $\ket{T_0}$ is given by:
\begin{align}
     \text{T}_1&= \frac{\sqrt{3}}{\widetilde{\delta b_z}},\\
     \text{T}_1^{eff}&= \frac{\sqrt{3}}{{\widetilde{\delta b_z}} \alpha}.
\end{align}

\section{Discussion}\label{Sec: Discussion}

\noindent Encoding qubits in the singlet-triplet states in DQD devices is an interesting approach since the spin angular momentum in direction $\hat{z}$ of those states is zero, implying resilience against hyperfine noise in the surrounding. Furthermore, one can theoretically isolate the $\ket{S}$ and $\ket{T_0}$ states to avoid transitions to other states such as $\ket{T_\pm}$. However, this is not experimentally easy since on an actual semiconductor heterostructure different sources of noise that break the $L_z$ and the spin-position symmetry exist, such as hyperfine and spin-orbit interactions~\cite{Burkard2002}. Noise reduces the coherence time of the state encoded in a qubit and results in erroneous results obtained after running a quantum algorithm.    

Since leakage usually comes from interactions with the environment, it contributes to incoherent errors and it is studied as a source of decoherence of the quantum state~\cite{LeakageErrors}. However, in this work we discussed that the presence of constant transverse magnetic fields in the system, which introduces leakage, has an impact on its dynamics, causing a shift in the quantum evolution operator and, hence, over or under rotations. We studied it under the assumption of constant pulses during the gate time, which is not far from the actual pulses implemented experimentally~\cite{Kawakami2014,T2noise}. In spite of that, it is worth noting the suitability of the Dyson series expansion employed here for time-dependent Hamiltonians compared to other perturbative techniques. While the SW transformation is a standard technique for deriving effective Hamiltonians, its generalizations require a time‑dependent generator and extra $-i \partial_t$ terms \cite{SWtime,reascos2025universalsolutionschriefferwolfftransformation}. In contrast, the Dyson formalism naturally accommodates the time-ordering required for describing pulse-driven evolution. This feature makes our analysis particularly robust for the time-dependent Hamiltonians, allowing for a direct extension to arbitrary pulse shapes beyond the rectangular profiles considered here.

As shown, leakage produces faster (or slower) rotations depending on the amplitude of the interaction and the energy gap between the computational basis and the other accessible states. External magnetic fields in QD architectures are typically implemented using micromagnets (see, for example, Fig. 1a in Ref.~\cite{micromagnet-fig}), which generate a total magnetic field of the form $\vec{B}_{tot}(\vec{r}) = \vec{B}_{ext} + \vec{B}(\vec{r})$~\cite{micromagnets,NoiseMicro,micromagnet}. Micromagnets produce a strong magnetic-field gradient along the $\hat{z}$ direction, aligned with the axis of the spin array, as well as weaker transverse field components that depend on the position $\vec{r}$ of each QD~\cite{SpinHEx,IBM-SPinHex}. Similar transverse magnetic fields may also arise in recently proposed QD-based architectures without micromagnets, where such fields can appear naturally~\cite{SpinHEx,IBM-SPinHex}. In addition, hyperfine interactions between spin qubits and the surrounding nuclear spin bath generate an effective Overhauser field with transverse components~\cite{HyperInt}. Although this field is not strictly constant, its evolution occurs on timescales much longer than typical gate operations, and it can therefore be treated as quasi-static during the gate time~\cite{Hyperfine}.

In this paper, we studied the ideal evolution of ST$_0$ spin qubits in DQD devices in the presence of leakage for single qubit operations. Notably, entangling gates between pairs of singlet-triplet qubits encoded in DQDs are realized by means of similar pulses, i.e. based on transverse magnetic fields and exchange interactions, and, thus, we expect similar results for those~\cite{Wang-CompositePulses}. Quantum algorithms require the application of a huge amount of very calibrated rotations over large amounts of qubits. For example, more than $10^{18}$ gates are needed for running Shor's algorithm in a fault-tolerant manner~\cite{Gidney2021howtofactorbit}. In this sense, although the time required to perform a single rotation is very short (on the order of nanoseconds), the coherence time of spin qubits is also quite limited: $T_1$ is on the order of milliseconds, and 
$T_2$ on the order of tens of microseconds~\cite{QuantumDots,RevModPhysSiQ}; leading to incorrect results. We showed that having control over the external magnetic fields gives the possibility to reduce or to increment the required time to perform certain rotations without a hard impact on the fidelity of the gates. Even a very small reduction of the required time to perform a rotation would have an important impact on the time needed to run a long quantum algorithm. 

Additionally, in the paradigm of QEC is very important to characterize all the different possible noise sources so that they can effectively be treated. The QEC code threshold refers to the physical error rate below which certain code effectively reduces the probability of logical errors for certain noise model~\cite{threshold,threshold1,threshold2}. To compute this value a characterization of the different sources of noise is needed. Leakage is usually considered as a probability of measuring a qubit on a state outside the computational basis and, thus, can be treated as an erasure errors~\cite{LeakageErrors} which pose more severe challenges than coherent phase errors. Nevertheless, as we discuss in this work, in the weak leakage regime it is also a source of gate errors due to over- or under-rotations which influences a QEC process or a quantum algorithm. Coherent errors can be made into stochastic errors by means of Pauli twirling, so leakage could contribute by increasing the gate physical error rate as discussed in this work~\cite{TonDecoders}.

In contrast, QEM techniques try to reduce the noise during the execution of a quantum algorithm using classical post-processing algorithms~\cite{ZNE}. In the NISQ era, QEM is a promising post-processing technique to improve the precision of current quantum computers until fault-tolerant quantum computers are constructed. Many different QEM techniques exist, all of which require a good characterization of the noise. For example, ZNE requires the existence of a tunable global noise source, i.e. a way to control all the noise rates of a system globally~\cite{Comment}. Usually, such global noise source can be obtained by controlling the time needed to run the algorithm, i.e. changing the time needed to perform each rotation. This noise amplification technique is referred to as pulse stretching~\cite{Kim2023,QEM1}. In this work, we showed that controlling leakage terms in the system enables the modification of the time required to perform a quantum rotation, its impact on the fidelity and on the non-unitary evolution of the system during the readout process, as we prove in sections \ref{Sec: fidelity} and \ref{Sec: Readout}, respectively. By using a magnetic transverse field we can not only tune the speed of the evolution, changing the duration over which the system is exposed to decoherence; but also change the effective coupling of the non-unitary operators during the readout process.

Additionally, if leakage is not properly accounted for, it may introduce uncertainty in the noise rates (i.e., the x-axis in extrapolation methods) used in QEM because it alters the unitary part of the evolution and the effective couplings. Although it may not be the best method to tune noise rates to use extrapolation-based error mitigation techniques, it has certainly an impact on their accuracy. A detailed analysis of these effects must take into account all noise sources of the system, their coupling with the qubit and the impact of leakage to each of them. We do not include this analysis in this work because it is beyond its scope.

\section{Conclusion} \label{Sec: Conclusion}

\noindent Qubits are encoded in quantum systems which, generally, have more than two accessible levels. Usually, the implementation of a qubit in physical devices, such as QDs, is done in such a way that the probability of leakage out of the computational subspace is very low. The main reason is to reduce the decoherence suffered by the system due these transitions. However, reducing those to zero is generally impossible, implying the emergence of leakage errors. 

In this work, we discussed how leakage has an impact on the  evolution of the system when an external field to perform a rotation is applied, producing over- or under-rotations. The change of the quantum evolution operator depends on the amplitudes of the interactions, the desired rotation and the energy gap between the energy levels of the system. Even if the difference is small, coherent errors accumulate for deep circuits resulting in ultimate failure. This result is important to understand the evolution of systems with more than two accessible levels with transitions among them induced by means of noise or gate control. In addition to this result, we showed that having control over the leakage terms not only allows to change the time required to perform a rotation but also it has an impact on the effective coupling of the lindbladian operators of the non-unitary evolution. 

Then, we derived bounds on the gate fidelity in the presence of charge noise, magnetic-field fluctuations, and weak transverse fields. We find that the fidelity is only weakly affected, owing to the large energy gap separating the computational states from higher-energy levels compared to the noise-induced couplings within the computational subspace. Nevertheless, leakage induces a small systematic shift that leads to coherent errors, underscoring the importance of accounting for leakage effects during gate design and calibration in realistic devices.

In addition, experimentally verifying the conclusions obtained would be important to confirm these results. As mentioned, we assume pulses that are easily implementable in experiment, so implementing experiments would not require exotic pulse generators. The phase shift produced by leakage might be related with some observed dephasing decoherence observed in GaAs singlet-triplet qubit~\cite{Hyperfine}, with the observed Rabi frequencies observed in~\cite{Holes1, Holes2} using holes spin qubits and with the spin-orbit signatures observed in~\cite{SOI}. The most challenging part would probably be the use of transverse magnetic fields that tune the leakage terms, but we consider that this can be handled and it might naturally appear in recently proposed architectures based on QDs, as we said above. As a result, a better understanding of the effect of leakage on the evolution of quantum systems would be obtained, which is critical to make practical quantum computations.

Finally, leakage errors are very detrimental for QEC, as seen for superconducting qubits~\cite{leakageqec1}. Simulating the impact of leakage in QEC codes can be challenging, but simulating their performance under coherent errors has been tackled efficiently~\cite{bravyiCoherent}. As a result, it would be interesting to determine if our approximations for the impact of leakage as coherent errors are accurate to simulate QEC systems. This would be very relevant to consider realistic noise in QEC architectures based on spin qubits, especially the ones making use of singlet-triplet qubits~\cite{spinQEC,SpinHEx,LDSTqec,2Narray}.

\section*{Acknowledgements}

We thank other members of the Hitachi-Cambridge Laboratory and the Quantum Information Group at Tecnun for their support and many useful discussions. This work was partially supported  by the Spanish Ministry of Science and Innovation through the project ``Few-qubit quantum hardware, algorithms and codes, on photonic and solidstate systems'' (PLEC2021-008251) and by the Diputación Foral de Gipuzkoa through the ``Biased quantum error mitigation and applications of quantum computing to gene regulatory networks'' project (2024-QUAN-000020). J.E.M. is funded by the Spanish Ministry of Science, Innovation and Universities through a Jose Castillejo mobility grant for his stay at the Cavendish Laboratory of the University of Cambridge.

\bibliographystyle{IEEEtran}
\bibliography{Bibliography}

\clearpage
\appendix

\section{Spin Qubits in Quantum Dots}\label{SecA: Spin Qubits}

\noindent QDs are created on semiconductors interfaces, where a small empty region is created in the 2 dimensional electron gas (2DEG) by applying an external potential $V_\text{ext}$. There are many heterostructures used to generate a 2DEG, where QDs can be created, such as Gallium arsenide (GaAs) and aluminum gallium arsenide (AlGaAs)\cite{doi:10.1126/science.278.5344.1792,LFedichkin_2000,li2001inas}; Silicon-Silicon Dioxide (Si/SiO$_2$) or silicon Germanium (SiGe)\cite{PhysRevB.82.155312,RevModPhysSiQ}. Nowadays, heterostructures based on purified silicon ($^{28}$Si) are taking a lot of attention because its spinless nucleus, i.e. there is not hyperfine interaction between them and an electron inside a QD~\cite{LongCoherenceT,Morton2011}. 

By controlling the external potential $V_\text{ext}$, the drain potential $V_d$ and the source potential $V_s$, which are the potentials which control the barrier between the QD and the 2DEG; we can change the charge energy to load electrons inside the QD~\cite{ChargeEnergy}. A qubit is usually encoded on a QD charged with a single electron, while an external magnetic field is applied to split the degenerate energies of the spin states $\ket{\uparrow} \equiv \ket{1}$ and $\ket{\downarrow} \equiv \ket{0}$ due to the Zeeman interaction, forming the computational basis. The Zeeman interaction Hamiltonian $\text{H}_Z(t)$ is given by:

\begin{align} \label{EqA: Zeeman Hamiltonian}
    \text{H}_Z(t) =  g \mu_B \vec{B}(t) \cdot \vec{s},
\end{align}
where g is the gyromagnetic ratio of the electron, $\mu_B$ is the Bohr magneton, $\vec{B}$ is the external magnetic field and $\cdot$ refers to the dot product. $\vec{s}= \frac12 \vec{\sigma}$ is the spin operator vector, with $\vec{\sigma}=\{\sigma_x, \sigma_y, \sigma_z\}$ the vector of Pauli matrices. Those matrices correspond to the generators of the SU(2) Lie algebra. To generate the qubit, we apply a constant external magnetic field in the $\hat{z}$ direction, i.e. $\vec{B}=(0,0,B_z)$. To control the state of the system, i.e. to apply quantum gates to a single qubit, external time-varying magnetic fields are applied in the direction we want to rotate the qubit, $\text{H}_C(\vec{B}(t))$. The total Hamiltonian of the system, accounting for both the qubit dynamics and the external controls, is given by:
\begin{align}
    \text{H}_{QD} + \text{H}_C(t) = b_z \sigma_z + \frac{1}{2} g \mu_B \vec{B}(t)\cdot \vec{\sigma}.
\end{align}

\noindent As a result, the rotation operator is defined in Eq.~\ref{Eq: BasicRot},  with a rotation angle $\theta(\lambda_x,\tau_{G_x})$, which depends on the amplitude of the interaction $\lambda_i = b_i$ and the gate time $\tau_{G_i}$, $\forall i \in \{x,y,z\}$.

Furthermore, the exchange interaction between electrons located in different QDs with the Zeeman interaction allows to perform multiqubit (entangling) gates. The coupling can be controlled by means of the \textit{voltage} detuning between their respective potentials $\varepsilon = V_{\text{Ext}_1}-V_{\text{Ext}_2}$, where $V_{\text{Ext}_i}$ is the potential applied to the i$^\text{th}$ QD and its sign is a function of the magnetic properties of the semiconductor: ferromagnetic or antiferromagnetic~\cite{T2noise}. For the present manuscript, we assume positive sign for $ \text{J}_\text{exc} [\varepsilon(t)]$ since the study is independent of its value, see Appendix~\ref{SecA: symmetry}. Consequently, the total Hamiltonian to perform entangling gates is:

\begin{align} \label{EqA: Exchange Hamiltonian}
    \text{H}_C(t) &= \text{H}_{exc}(t) + \sum_{j=1,2} \text{H}_{Z,j}(t) \\
    &=   \text{J}_\text{exc}[\varepsilon(t)] \vec{s}_1 \cdot \vec{s}_2 +  \sum_{i=1,2} g_i \mu_B \vec{B}(r_i,t) \cdot \vec{s}_i,
\end{align}

where vector $\vec{s}_i= 1/2 \vec{\sigma}_i$ contains the spin operators of the electron in the i$^\text{th}$ QD, $\pm  \text{J}_\text{exc}[\varepsilon(t)]$ is the exchange coupling strength between them and $\vec{B}_i(r_i,t)$ is the magnetic field in the position of the i$^{th}$ QD. Finally, qubit readout is realized by means of the so-called spin-to-charge connection technique~\cite{Elzerman2004}. 

\subsection{Spin Qubits in Double Quantum Dots}

\noindent Spin qubits in QD infrastructures can be encoded in DQD systems with two electrons. Although two QDs are needed to encode a single qubit, this approach is advantageous because it can be resilient to certain sources of decoherence, such as hyperfine noise, because the SU(2) $\otimes$ SU(2) Hilbert space has a subspace without spin angular momentum in $\hat{z}$, see Appendix~\ref{SecA: symmetry}. As a result, these systems exhibit longer coherence times, $T_1$ and $T_2$~\cite{FreeSubspaceST0}. The eigenstates of a system formed by $2$ electrons are the singlet $\ket{S}$ and the triplet states ($\ket{T_0}$ and $\ket{T_\pm}$), which belong to the U(1) and SU(3) subspaces, respectively. Those states are defined as

\begin{align}
     \ket{S}&= \frac{\ket{\uparrow \downarrow}- \ket{\downarrow \uparrow}}{\sqrt{2}}, \\
     \ket{T_+} &= \ket{\uparrow \uparrow}, \\
     \ket{T_0} &= \frac{\ket{\uparrow \downarrow}+ \ket{\downarrow \uparrow}}{\sqrt{2}}, \\
     \ket{T_-} &= \ket{\downarrow \downarrow}. 
\end{align}

\noindent The singlet state is anti-symmetric in its spin wave function and symmetric in its position wave function, allowing both electrons to occupy the same QD. In contrast, triplet states have a symmetric wave function in the spin basis and the electrons can not occupy the same QD due to Pauli exclusion principle. Along this document, we study qubits encoded in a DQDs for the case that a single electron is loaded in each QD, i.e there is no hopping from one QD to the other. The interactions we are interested in are the exchange interaction, $\text{H}_\text{exc}$ (Eq.~\ref{EqA: Exchange Hamiltonian}) and the Zeeman interaction, $\text{H}_\text{Z}$ (Eq.~\ref{Eq: Zeeman Hamiltonian}). Therefore, the Hamiltonian of a DQD system in the  $\{\ket{S}$, $\ket{T_{0}}$, $\ket{T_+}$, $\ket{T_-}\}$ basis is given by:
\begin{widetext}
\begin{align} \label{EqA: Hamil6x6}
\text{H}&= \text{H}_\text{Z} + \text{H}_\text{exc}      =\begin{pmatrix}
         -\frac18  \text{J}_\text{exc}[\varepsilon] & \frac{1}{2} g \mu_B \delta B_z & -\frac{1}{2\sqrt{2}} g \mu_B [\delta B_x+i\delta B_y] & \frac{1}{2\sqrt{2}} g \mu_B [\delta B_x-i\delta B_y] \\
         \frac{1}{2} g \mu_B \delta B_{z} & \frac18  \text{J}_\text{exc}[\varepsilon] & \frac{1}{2\sqrt{2}} g \mu_B [B_x+iB_y] & \frac{1}{2\sqrt{2}} g \mu_B [B_x- i  B_y] \\
        -\frac{1}{2\sqrt{2}} g \mu_B [\delta B_x-i\delta B_y] & \frac{1}{2\sqrt{2}} g \mu_B [B_x-iB_y] & \frac18  \text{J}_\text{exc}[\varepsilon] + \frac12 g\mu_B B_z & 0 \\
        \frac{1}{2\sqrt{2}} g \mu_B [\delta B_x+i\delta B_y] & \frac{1}{2\sqrt{2}} g \mu_B [B_x+i B_y] & 0 &  \frac18  \text{J}_\text{exc}[\varepsilon] - \frac12 g\mu_B B_z
    \end{pmatrix},
\end{align}
\end{widetext}
where the explicit dependence on $t$ has been removed to simplify the notation; $B_;x = B_{x_1} + B_{x_2}$, $B_y = B_{y_1} + B_{y_2}$, $B_z = B_{z_1} + B_{z_2}$, $\delta B_x = (B_{x_1} - B_{x_2})$, $\delta B_y = (B_{y_1} - B_{y_2})$ and $\delta B_z = (B_{z_1} - B_{z_2})$. This Hamiltonian can be deduced from the Hamiltonian of the different interactions and from symmetry arguments (see Appendix~\ref{SecA: symmetry}). There are different ways to encode a logical qubit in this system, such as ST$_+$ or flip-flop qubits (see Appendix~\ref{SecA: Other encoded qubits}). In this work, we focus on ST$_0$ qubits, where the computational basis are formed by the states $\ket{S} \equiv \ket{0}$ and $\ket{T_0}\equiv \ket{1}$.

\subsection{Singlet-Neutral triplet Qubits} \label{Sec: ST0qubits}

\noindent The ST$_0$ qubit is encoded in the singlet $\ket{S}$ and neutral triplet $\ket{T_0}$ eigenstates of a DQD. To get this encoding, the Zeeman split of the energy levels should be $B_z >>  \text{J}_\text{exc}$ in order to make the energy gap between $\ket{T_\pm}$ states and the computational basis ($\ket{S}$ and  $\ket{T_0}$) large enough. In this way, possible leakage from the computational subspace to the other accessible triplet states is reduced. We can write the effective Hamiltonian of the ST$_0$ from the Hamiltonian given in Eq.~\ref{Eq: Hamil6x6}. The energy levels defining the encoded qubit are the states $\ket{S}$ and $\ket{T_0}$ given by the Hamiltonian $\text{H}_{ST_0}$ given in Eq.~\ref{Eq: ST0Hamil}. To perform a rotation around the $\hat{x}$ axis alone is not possible; instead, rotations around  a combination of $\hat{x}$ and $\hat{z}$ axis are feasible. The Hamiltonian of the interaction between the qubit and an external magnetic field $\delta B_z\left(\vec{r_1},\vec{r_2}\right)= B_{z}(\vec{r_1},t)-B_{z}(\vec{r_2},t)$ in the ST$_0$ subspace is: 

\begin{align} 
    \text{H}_\text{Z}(t) =g \mu_B \frac{\delta B_z(t)}{2} \sigma_x  = g \mu_B \begin{pmatrix}
        0 & \delta B_z(t)/2 \\
        \delta B_z(t)/2 & 0
    \end{pmatrix}.
\end{align}

\noindent Then, rotations around $\hat{z}$ are achieved by manipulating the detuning to change the exchange coupling $ \text{J}_\text{exc}[\varepsilon(t)]$ and around $\hat{x}\hat{z}$ by manipulating the external magnetic field $\delta B_z$. Respectively, the Hamiltonians are given by:
\begin{align}
   \text{H}_{R_Z}(t) &= \frac{ \text{J}_\text{exc}[\varepsilon(t)]}{8} \sigma_z  =  \begin{pmatrix}
        \text{J}_\text{exc}[\varepsilon(t)]/8 & 0 \\
       0 &  -\text{J}_\text{exc}[\varepsilon(t)]/8
   \end{pmatrix}\label{EqA: Z rot} \\
   \text{H}_{R_{XZ}}(t) &= \delta b_z(t) \sigma_x + \frac{ \text{J}_\text{exc}[\varepsilon(t)]}{8} \sigma_z \nonumber \\
    &= \begin{pmatrix}
        \frac{ \text{J}_\text{exc}[\varepsilon(t)]}{8} &  \delta b_z(t) \\
        \delta b_z(t) & \frac{ -\text{J}_\text{exc}[\varepsilon(t)]}{8}
    \end{pmatrix}  \label{EqA: XZ Rot}
\end{align}

\begin{figure}
    \centering
    \includegraphics[width=0.9\textwidth]{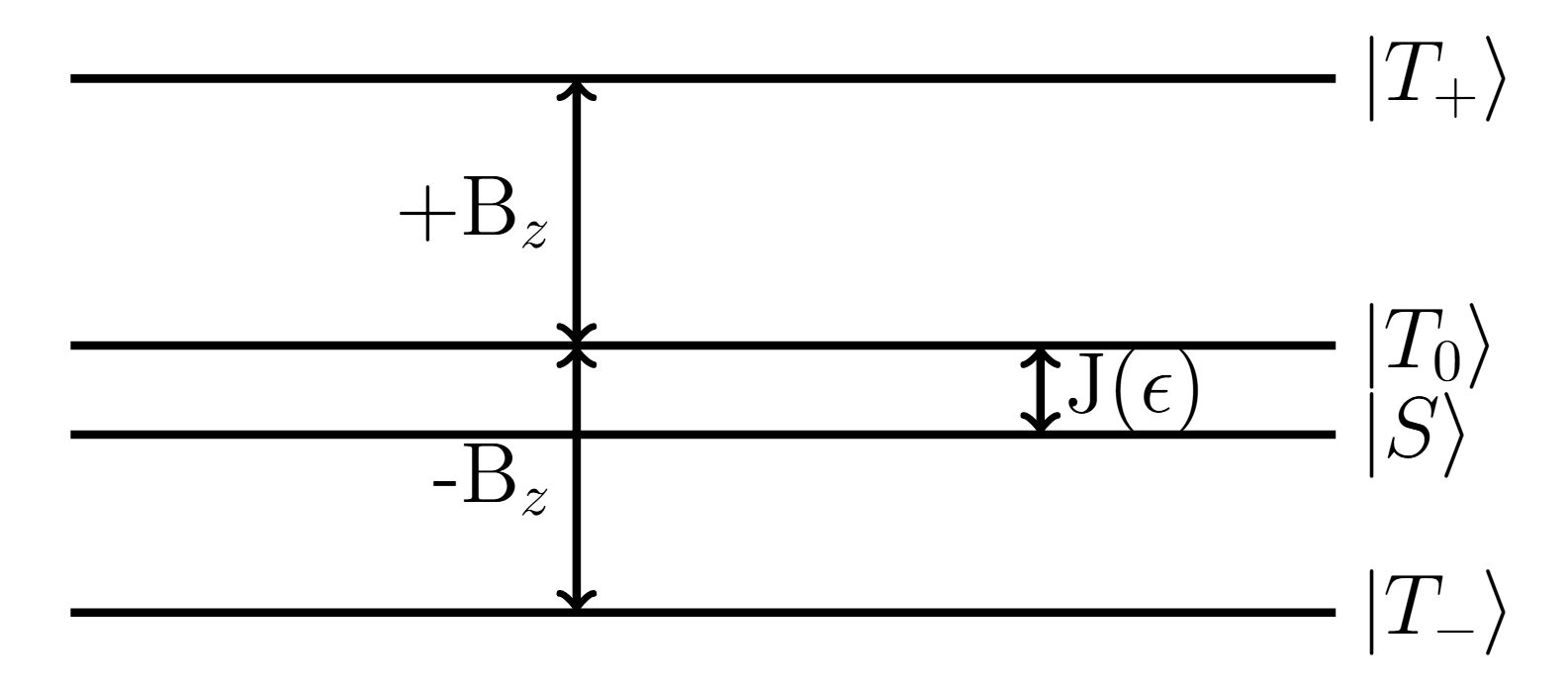}
    \caption{Schematic representation of the energy levels of a DQD system with one electron in each QD. The Lie algebra is given by SU(2) $\otimes$ SU(2) $=$ U(1) $\oplus$ SU(3), i.e. the singlet and the triplets states. The Hamiltonian of the system is given in Eq.\ref{Eq: Hamil6x6} with $\vec{B}= (0,0,B_z)$ and $\delta \vec{B} = (0,0,0)$} 
    \label{Fig: STO diagram1}
\end{figure}
\noindent Then the rotation operators are given by:

\begin{align}
    &R(\text{H}(t),\tau_G) = \mathcal{T} \left\{ \text{exp}\left(\frac{-i}{\hbar} \int_0^{\tau_G} dt  \text{H}(t) \right) \right\} \\
    &= \text{exp}\left(\frac{-i}{\hbar} \int_0^{\tau_G} dt \hspace{0.1cm} g \mu_B\frac{\delta B_z(t)}{2} \sigma_x + \frac{ \text{J}_\text{exc}[\varepsilon(t)]}{8} \sigma_z \right) \\ 
    &=\text{exp}\left(\frac{-i}{\hbar}\left(g \mu_B\frac{\delta B_z}{2} \sigma_x + \frac{ \text{J}_\text{exc}[\varepsilon]}{8} \sigma_z\right) \tau_G \right), \label{EqA: RotationGen}
\end{align}
where $\mathcal{T} \left\{ \cdot \right\}$ is the time ordering operator. Under the assumption of square pulses the Hamiltonian commuted with itself  $[H(t_1),H(t_2)]=0$ and  $ \text{J}_\text{exc}[\varepsilon(t)]=  \text{J}_\text{exc}[\varepsilon]$ and $\delta B_z(t) = \delta B_z$ remain constants $\forall t_i \in [0,\tau_G]$. We define the angles as:
\begin{align}
    \theta_x &=\tau_G \lambda_x= \tau_G g \mu_B\delta B_z, \\
    \theta_z &=\tau_G \lambda_z = \tau_G \frac{ \text{J}_\text{exc}[\varepsilon]}{4}.
\end{align}

\noindent Then, the rotation operator is defined as:
\begin{align}\label{EqA: Rotation}
    R_{\hat{r}} (\theta_x + \theta_z)= \text{exp}\left(\frac{-i}{\hbar} \frac{\theta_x\sigma_x + \theta_z\sigma_z}{2}\right),
\end{align}
where the rotation axis is:
\begin{align}
\hat{r}=\frac{g \mu_B\frac{\delta B_z}{2} \hat{x} + \frac{ \text{J}_\text{exc}[\varepsilon]}{8} \hat{z}}{\sqrt{\left(\frac{ \text{J}_\text{exc}[\varepsilon]}{8}\right)^2  +\left(\frac{ g \mu_B\delta B_z}{2}\right)^2 }}.
\end{align}

\noindent These are the main interactions involving ST$_0$ qubits, any other rotation or entangling gate is formed by a combination of these interactions with different values of $ \text{J}_{\text{exc};ij}[\varepsilon]$ and $\delta B_{z; ij}$ where $i$ and $j$ refer to different QDs~\cite{Wang-CompositePulses}.

\subsection{Exchange interaction}
¡Recall that the Hamiltonian of the exchange interaction is:
\begin{align}
    \text{H}_\text{exc}=  \text{J}_\text{exc} \vec{s}_1 \cdot \vec{s}_2.
\end{align}

\noindent The expected energy due to this interaction for the different states are:
\begin{align}
    \bra{S}\text{H}_\text{exc}\ket{S}&= 0, \\
    \bra{T_0}\text{H}_\text{exc}\ket{T_0}&=\frac14  \text{J}_\text{exc}, \\
    \bra{T_+}\text{H}_\text{exc}\ket{T_+}&=\frac14  \text{J}_\text{exc} , \\
    \bra{T_0}\text{H}_\text{exc}\ket{T_-}&=\frac14  \text{J}_\text{exc} .
\end{align}

\subsection{Zeeman interaction}
\noindent Recall that the Hamiltonian of the Zeeman interaction is:
\begin{align}
     \text{H}_\text{Z}(B,\vec{r}) = g \mu_B \vec{B}(\vec{r},t)\cdot(\vec{S}_1(\vec{r}_1) + \vec{S}_2(\vec{r}_2),
\end{align}

where $\vec{B}(\vec{r})$ is the external magnetic field which depends on the position $\vec{r}_i$ of the i$^{th}$ QD, $\mu_B$ is the Bohr magneton and $g$ is the gyromagnetic ratio. The interaction of the singlet state $\ket{S}$ with an external magnetic field is:
\begin{align}
    \text{H}_\text{Z}(B_x) \ket{S} &=  g \mu_B [B_{x_1}(X \otimes I) + B_{x_2}(I \otimes X)] \ket{S}\\
    &= \frac12 g \mu_B \left[ \frac{B_{x_1}\left(\ket{T_-}-\ket{T_+}\right)}{\sqrt{2}} + \frac{B_{x_2}\left(\ket{T_+}-\ket{T_-}\right)}{\sqrt{2}}\right]\\
    &=\frac{1}{2 \sqrt{2}}g \mu_B (B_{x_1}-B_{x_2}) (\ket{T_-} - \ket{T_+}), \\
    \text{H}_\text{Z}(B_y) \ket{S} &=   g \mu_B [B_{y_1}(Y \otimes I) + B_{y_2}(I \otimes Y)] \ket{S}\\
    &= \frac12 g \mu_B \left[ \frac{iB_{y_1}\left(\ket{T_-}+\ket{T_+}\right)}{\sqrt{2}} - \frac{iB_{y_2}\left(\ket{T_+}+\ket{T_-}\right)}{\sqrt{2}}\right]\\
    &=\frac{i}{2 \sqrt{2}}g \mu_B (B_{y_1}-B_{y_2}) (\ket{T_-} + \ket{T_+}), \\
    \text{H}_\text{Z}(B_z) \ket{S} &=  g \mu_B [B_{z_1}(Z \otimes I) + B_{z_2}(I \otimes Z)] \ket{S}\\
    &= \frac12 g \mu_B \left[ \frac{B_{z_1} \ket{T_0}}{\sqrt{2}} - \frac{B_{z_2}\ket{T_0}}{\sqrt{2}}\right]\\
    &=\frac{1}{2 }g \mu_B (B_{z_1}-B_{z_2}) \ket{T_0}.
\end{align}

\noindent The interaction of the neutral triplet state $\ket{T_0}$ with an external magnetic field is:

\begin{align}
    \text{H}_\text{Z}(B_x) \ket{T_0} &=  g \mu_B [B_{x_1}(X \otimes I) + B_{x_2}(I \otimes X)] \ket{T_0}\\
    &= \frac12 g \mu_B \left[ \frac{B_{x_1}\left(\ket{T_-}+\ket{T_+}\right)}{\sqrt{2}} + \frac{B_{x_2}\left(\ket{T_+}+\ket{T_-}\right)}{\sqrt{2}}\right] \\
    &=\frac{1}{2\sqrt{2}}g \mu_B (B_{x_1}+B_{x_2}) (\ket{T_-} + \ket{T_+}), \\
    \text{H}_\text{Z}(B_y) \ket{T_0} &=   g \mu_B [B_{y_1}(Y \otimes I) + B_{y_2}(I \otimes Y)] \ket{T_0}\\
    &= \frac12 g \mu_B \left[ \frac{iB_{y_1}\left(\ket{T_-}-\ket{T_+}\right)}{\sqrt{2}} + \frac{iB_{y_2}\left(\ket{T_-}-\ket{T_+}\right)}{\sqrt{2}}\right] \\
    &=\frac{i}{2 \sqrt{2}}g \mu_B (B_{y_1}+B_{y_2}) (\ket{T_-} - \ket{T_+}), \\
    \text{H}_\text{Z}(B_z) \ket{T_0} &=  g \mu_B [B_{z_1}(Z \otimes I) + B_{z_2}(I \otimes Z)] \ket{T_0}\\
    &= \frac12 g \mu_B \left[ \frac{B_{z_1} \ket{S}}{\sqrt{2}} - \frac{B_{z_2}\ket{S}}{\sqrt{2}}\right]\\
    &=\frac{1}{2 }g \mu_B (B_{z_1}-B_{z_2}) \ket{S}.
\end{align}

\noindent The interaction of the positive triplet state $\ket{T_+}$ with an external magnetic field is:

\begin{align}
    \text{H}_\text{Z}(B_x) \ket{T_+} &=  g \mu_B [B_{x_1}(X \otimes I) + B_{x_2}(I \otimes X)] \ket{T_+}\\
    &= \frac12 g \mu_B \left[ \frac{B_{x_1}\left(\ket{T_0}-\ket{S}\right)}{\sqrt{2}} + \frac{B_{x_2}\left(\ket{T_0}+\ket{S}\right)}{\sqrt{2}}\right] \\
    &=\frac{1}{2\sqrt{2}}g \mu_B [(B_{x_1}+B_{x_2}) \ket{T_0} -(B_{x_1}-B_{x_2}) \ket{S} ], \\
    \text{H}_\text{Z}(B_y) \ket{T_+} &=   g \mu_B [B_{y_1}(Y \otimes I) + B_{y_2}(I \otimes Y)] \ket{T_+}\\
    &= \frac12 g \mu_B \left[ \frac{iB_{y_1}\left(\ket{T_0}-\ket{S}\right)}{\sqrt{2}} + \frac{iB_{y_2}\left(\ket{T_0}+\ket{S}\right)}{\sqrt{2}}\right] \\
    &=\frac{i}{2\sqrt{2}}g \mu_B [(B_{y_1}+B_{y_2}) \ket{T_0} -(B_{y_1}-B_{y_2}) \ket{S} ], \\
    \text{H}_\text{Z}(B_z) \ket{T_+} &=  g \mu_B [B_{z_1}(Z \otimes I) + B_{z_2}(I \otimes Z)] \ket{T_+}\\
    &=\frac{1}{2 }g \mu_B (B_{z_1}+B_{z_2}) \ket{T_+}.
\end{align}

\noindent The interaction of the negative triplet state $\ket{T_-}$ with an external magnetic field is:
\begin{align}
    \text{H}_\text{Z}(B_x) \ket{T_-} &=  g \mu_B [B_{x_1}(X \otimes I) + B_{x_2}(I \otimes X)] \ket{T_-}\\
    &= \frac12 g \mu_B \left[ \frac{B_{x_1}\left(\ket{T_0}+\ket{S}\right)}{\sqrt{2}} + \frac{B_{x_2}\left(\ket{T_0}-\ket{S}\right)}{\sqrt{2}}\right] \\
    &=\frac{1}{2\sqrt{2}}g \mu_B [(B_{x_1}+B_{x_2}) \ket{T_0} +(B_{x_1}-B_{x_2}) \ket{S} ], \\
    \text{H}_\text{Z}(B_y) \ket{T_-} &=   g \mu_B [B_{y_1}(Y \otimes I) + B_{y_2}(I \otimes Y)] \ket{T_-}\\
    &= \frac12 g \mu_B \left[ \frac{-iB_{y_1}\left(\ket{T_0}+\ket{S}\right)}{\sqrt{2}} + \frac{-iB_{y_2}\left(\ket{T_0}-\ket{S}\right)}{\sqrt{2}}\right] \\
    &=\frac{-i}{2\sqrt{2}}g \mu_B [(B_{y_1}+B_{y_2}) \ket{T_0} +(B_{y_1}-B_{y_2}) \ket{S} ], \\
    \text{H}_\text{Z}(B_z) \ket{T_-} &=  g \mu_B [B_{z_1}(Z \otimes I) + B_{z_2}(I \otimes Z)] \ket{T_-}\\
    &=\frac{-1}{2 }g \mu_B (B_{z_1}+B_{z_2}) \ket{T_-}.
\end{align}

\noindent Then the non-zero transitions from the singlet state $\ket{S}$ due to this interactions are:
\begin{align}
    \bra{T_+} \text{H}_\text{Z}(B_x) \ket{S} &= \frac{-1}{2 \sqrt{2}}g \mu_B (B_{x_1} - B_{x_2}), \\ 
    \bra{T_-} \text{H}_\text{Z}(B_x) \ket{S} &= \frac{1}{2 \sqrt{2}}g \mu_B (B_{x_1} - B_{x_2}), \\ 
    \bra{T_+} \text{H}_\text{Z}(B_y) \ket{S} &= \frac{i}{2 \sqrt{2}}g \mu_B (B_{y_1} - B_{y_2}), \\ 
    \bra{T_-} \text{H}_\text{Z}(B_y) \ket{S} &= \frac{i}{2 \sqrt{2}}g \mu_B (B_{y_1} - B_{y_2}), \\ 
    \bra{T_0} \text{H}_\text{Z}(B_z) \ket{S} &= \frac{1}{2}g \mu_B (B_{z_1} - B_{z_2}).
\end{align}

\noindent The transitions from the neutral triplet state $\ket{T_0}$ are:
\begin{align}
    \bra{T_+} \text{H}_\text{Z}(B_x) \ket{T_0} &= \frac{1}{2 \sqrt{2}}g \mu_B (B_{x_1} + B_{x_2}), \\ 
    \bra{T_-} \text{H}_\text{Z}(B_x) \ket{T_0} &= \frac{1}{2 \sqrt{2}}g \mu_B (B_{x_1} + B_{x_2}), \\ 
    \bra{T_+} \text{H}_\text{Z}(B_y) \ket{T_0} &= \frac{-i}{2 \sqrt{2}}g \mu_B (B_{y_1} + B_{y_2}), \\ 
    \bra{T_-} \text{H}_\text{Z}(B_y) \ket{T_0} &= \frac{i}{2 \sqrt{2}}g \mu_B (B_{y_1} + B_{y_2}), \\ 
    \bra{S} \text{H}_\text{Z}(B_z) \ket{T_0} &= \frac{1}{2}g \mu_B (B_{z_1} - B_{z_2}).
\end{align}
\noindent The transitions from the neutral triplet state $\ket{T_+}$ are:
\begin{align}
     \bra{S} \text{H}_\text{Z}(B_x) \ket{T_+} &= \frac{-1}{2 \sqrt{2}}g \mu_B (B_{x_1} - B_{x_2}), \\ 
    \bra{T_0} \text{H}_\text{Z}(B_x) \ket{T_+} &= \frac{1}{2 \sqrt{2}}g \mu_B (B_{x_1} + B_{x_2}), \\ 
    \bra{S} \text{H}_\text{Z}(B_y) \ket{T_+} &= \frac{-i}{2 \sqrt{2}}g \mu_B (B_{y_1} - B_{y_2}), \\ 
    \bra{T_0} \text{H}_\text{Z}(B_y) \ket{T_+} &= \frac{i}{2 \sqrt{2}}g \mu_B (B_{y_1} + B_{y_2}), \\ 
    \bra{T_+} \text{H}_\text{Z}(B_z) \ket{T_+} &= \frac{1}{2}g \mu_B (B_{z_1} + B_{z_2}).
\end{align}

\noindent The transitions from the neutral triplet state $\ket{T_-}$ are:
\begin{align}
    \bra{S} \text{H}_\text{Z}(B_x) \ket{T_-} &= \frac{1}{2 \sqrt{2}}g \mu_B (B_{x_1} - B_{x_2}), \\ 
    \bra{T_0} \text{H}_\text{Z}(B_x) \ket{T_-} &= \frac{1}{2 \sqrt{2}}g \mu_B (B_{x_1} + B_{x_2}), \\ 
    \bra{S} \text{H}_\text{Z}(B_y) \ket{T_-} &= \frac{-i}{2 \sqrt{2}}g \mu_B (B_{y_1} - B_{y_2}), \\ 
    \bra{T_0} \text{H}_\text{Z}(B_y) \ket{T_-} &= \frac{-i}{2 \sqrt{2}}g \mu_B (B_{y_1} + B_{y_2}), \\ 
    \bra{T_-} \text{H}_\text{Z}(B_z) \ket{T_-} &= \frac{-1}{2}g \mu_B (B_{z_1} + B_{z_2}).
\end{align}

\noindent Finally, the Hamiltonian of a DQD with two electrons with an exchange interaction among them and external magnetic field can be expressed in the  $\{\ket{S}$, $\ket{T_{0}}$, $\ket{T_+}$, $\ket{T_-}\}$ basis as:
\begin{widetext}
\begin{align} \label{EqA: DQD Hamiltonian}
    \text{H}_{DQD}&=\begin{pmatrix}
        0 & \frac{1}{2} g \mu_B \delta B_z & -\frac{1}{2\sqrt{2}} g \mu_B [\delta B_x+i\delta B_y] & \frac{1}{2\sqrt{2}} g \mu_B [\delta B_x-i\delta B_y] \\
         \frac{1}{2} g \mu_B \delta B_{z} & \frac14  \text{J}_\text{exc} & \frac{1}{2\sqrt{2}} g \mu_B [B_x+iB_y] & \frac{1}{2\sqrt{2}} g \mu_B [B_x- i  B_y] \\
        -\frac{1}{2\sqrt{2}} g \mu_B [\delta B_x-i\delta B_y] & \frac{1}{2\sqrt{2}} g \mu_B [B_x-iB_y] & \frac14  \text{J}_\text{exc} + \frac12 g\mu_B B_z & 0 \\
        \frac{1}{2\sqrt{2}} g \mu_B [\delta B_x+i\delta B_y] & \frac{1}{2\sqrt{2}} g \mu_B [B_x+i B_y] & 0 &  \frac14  \text{J}_\text{exc} - \frac12 g\mu_B B_z
    \end{pmatrix},
\end{align}
\end{widetext}

where $B_x = B_{x_1} + B_{x_1}$, $B_y = B_{y_1} + B_{y_1}$, $B_z = B_{z_1} + B_{z_1}$, $\delta B_x = (B_{x_1} - B_{x_2})$, $\delta B_y = (B_{y_1} - B_{y_2})$ and $\delta B_z = (B_{z_1} - B_{z_2})$.

\section{SU(2) $\otimes$ SU(2) symmetry } \label{SecA: symmetry}
\noindent The algebra of the direct product of two SU(2) spaces is equal to the direct sum of two subspaces:
\begin{align}
    SU(2) \otimes SU(2) = U(1) \oplus SU(3).
\end{align}

\noindent The symmetry $U(1)$ is associated to the singlet, with a symmetric wavefunction in position and asymmetric wavefunction in spin. The subspace of the singlet $\ket{S}$does not have spin angular momentum in direction $\hat{z}$:
\begin{align}
     \bra{S} \hat{S}_z \ket{S} = 0,
\end{align}

where $\hat{S}_z$ is the spin angular momentum in direction $\hat{z}$ operator. The symmetry group $SU(3)$ is associated to the triplets, with symmetric wavefunction in spin and a asymmetric wavefunction in position. This subspace has a total spin angular momentum equal to 1 and:
 \begin{align}
    \bra{T_-} \hat{S}_z \ket{T_-} &= -1, \\
    \bra{T_0} \hat{S}_z \ket{T_0} &= 0,  \\
    \bra{T_+} \hat{S}_z \ket{T_+} &= 1 .   
\end{align}
 
\noindent Then, we can define the Hamiltonian of the system as: 
\begin{align}
    \text{H} =U(1) \oplus SU(3)= \left(
\begin{array}{c|c}
    U(1) & 0 \\
    \hline  \\
     0 & SU(3) \\ 
      & 
\end{array}
\right).
\end{align}

\noindent U(1) only has 1 generator and the generators of $SU(3)$ are given by the Gell-Mann matrices:
\begin{align} \label{EqA: GellMann}
    \Lambda_1 &= \begin{pmatrix}
        0 & 1 & 0 \\
        1 & 0 & 0 \\
        0 & 0 & 0
    \end{pmatrix}, \nonumber \\
    \Lambda_2 &= \begin{pmatrix}
        0 & -i & 0 \\
        i & 0 & 0 \\
        0 & 0 & 0
    \end{pmatrix}, \nonumber\\
    \Lambda_3 &= \begin{pmatrix}
        1 & 0 & 0 \\
        0 & -1 & 0 \\
        0 & 0 & 0
    \end{pmatrix}, \nonumber \\
    \Lambda_4 &= \begin{pmatrix}
        0 & 0 & 1 \\
        0 & 0 & 0 \\
        1 & 0 & 0
    \end{pmatrix}, \nonumber\\
    \Lambda_5 &= \begin{pmatrix}
        0 & 0 & -i \\
        0 & 0 & 0 \\
        -i & 0 & 0
    \end{pmatrix}, \nonumber\\
    \Lambda_6 &= \begin{pmatrix}
        0 & 0 & 0 \\
        0 & 0 & 1 \\
        0 & 1 & 0
    \end{pmatrix}, \nonumber\\
    \Lambda_7 &= \begin{pmatrix}
        0 & 0 & 0 \\
        0 & 0 & -i \\
        0 & i & 0
    \end{pmatrix}, \nonumber\\
    \Lambda_8 &= \frac{1}{\sqrt{3}} \begin{pmatrix}
        1 & 0 & 0 \\
        0 & 1 & 0 \\
        0 & 0 & -2
    \end{pmatrix}.
\end{align}

\noindent The energy of each subspace is given by $\eta = diag\left(-1,1,1,1\right)$, which can be seen as the Minkowski metric of the system:
\begin{align} \label{EqA: HEnergy}
    \text{H}_{\eta}= \frac18  \text{J}_\text{exc} \eta = \frac18  \text{J}_\text{exc}\begin{pmatrix}
        -1 & 0 & 0 & 0 \\
         0 & 1 & 0 & 0 \\
         0 & 0 & 1 & 0 \\
         0 & 0 & 0 & 1     
    \end{pmatrix}.
\end{align}

\noindent Using the Gell-Mann matrices, we are able to reconstruct the Hamiltonian of the subspace related to the triplets. The external magnetic field in the direction $\hat{z}$ splits the space in 3 different states with different spin in $\hat{z}$ $S_z$, this interaction in related to $\{\Lambda_1,\Lambda_6\}$. The external magnetic fields in the directions $\hat{x}$ and $\hat{y}$ changes the spin by $\Delta S_z = \pm 1$, so they allow transitions between states with different spin separated by 1 and are related to the matrices $\{\Lambda_1,\Lambda_2,\Lambda_6,\Lambda_7\}$. There are no interactions allowing transition between states with $\Delta S_z = \pm 2$, so the Hamiltonian does not have terms related to $\{\Lambda_4,\Lambda_5\}$. Thus, the Hamiltonian in the basis $\{T_+, T_0, T_-\}$:
\begin{align} \label{EqA: HintT}
    \text{H}_{T_0,T_\pm} &= \frac12 g \mu_B \biggl( \frac{B_z}{2}\Lambda_3 + \frac{\sqrt{3}B_z}{2} \Lambda_8  \\
     &+ \frac{1}{\sqrt{2}} B_x \Lambda_1 + \frac{1}{\sqrt{2}} B_x \Lambda_6 + \frac{1}{\sqrt{2}} B_y \Lambda_2 + \frac{1}{\sqrt{2}} B_y \Lambda_7 \biggr) \\
    &= \frac12 g \mu_B \small\begin{pmatrix}
        B_z & \frac{1}{\sqrt{2}} \left(B_x -i B_y\right) & 0 \\
        \frac{1}{\sqrt{2}} \left(B_x +i B_y\right) & 0 & \frac{1}{\sqrt{2}} \left(B_x -i B_y\right) \\
        0 & \frac{1}{\sqrt{2}} \left(B_x +i B_y\right) & -B_z
    \end{pmatrix}.
\end{align}

\noindent Then, the Hamiltonian of the SU(2) $\otimes$ SU(2) system is given by the block diagonal matrix:
\begin{align}
    \text{H} &= \text{H}_{\eta} + \begin{pmatrix}
        0 & \\
         & \text{H}_{T_0,T_\pm}
    \end{pmatrix}.
\end{align}

\noindent In such systems we are able to break the symmetry by applying different magnetic fields to each QD, i.e. $B_{i_1} \neq B_{i_2}$ with $i \in \{x,y,z\}$. In this case, each subspace is not protected by this symmetry and transitions between them are possible. The matrix associated to that transitions are given by:
\begin{align} \label{EqA: GenSB}
    \Lambda'_1 &= \begin{pmatrix}
         0 & 1 & 0 & 0 \\
         1 & 0 & 0 & 0 \\
         0 & 0 & 0 & 0 \\
         0 & 0 & 0 & 0    
    \end{pmatrix} \nonumber \\
    \Lambda'_2 &= \begin{pmatrix}
         0 & -i & 0 & 0 \\
         i & 0 & 0 & 0 \\
         0 & 0 & 0 & 0 \\
         0 & 0 & 0 & 0    
    \end{pmatrix} \nonumber \\
    \Lambda'_3 &= \begin{pmatrix}
         0 & 0 & 1 & 0 \\
         0 & 0 & 0 & 0 \\
         1 & 0 & 0 & 0 \\
         0 & 0 & 0 & 0    
    \end{pmatrix} \nonumber  \\
    \Lambda'_4 &= \begin{pmatrix}
         0 & 0 & -i & 0 \\
         0 & 0 & 0 & 0 \\
         i & 0 & 0 & 0 \\
         0 & 0 & 0 & 0    
    \end{pmatrix} \nonumber  \\
    \Lambda'_5 &= \begin{pmatrix}
         0 & 0 & 0 & 1 \\
         0 & 0 & 0 & 0 \\
         0 & 0 & 0 & 0 \\
         1 & 0 & 0 & 0    
    \end{pmatrix} \nonumber \\
    \Lambda'_6 &= \begin{pmatrix}
         0 & 0 & 0 & -i \\
         0 & 0 & 0 & 0 \\
         0 & 0 & 0 & 0 \\
         i & 0 & 0 & 0    
    \end{pmatrix}.
\end{align}

\noindent A different magnetic field in each QD breaks the spin-position symmetry, then the interactions between the electrons and the external magnetic field in this situation allows transitions between subspaces U(1) and SU(3). An external field with $\delta B_z \neq 0$ does not give a angular momentum in the $\hat{z}$ direction, but it allows transitions between the states $\ket{S}$ and $\ket{T_0}$. In the situation of $\delta B_x, \delta B_y \neq 0$, transitions between states in different subspaces and with different angular momentum are allowed. The Hamiltonian of the interactions which break the symmetry is given by:
\begin{align} \label{EqA: HintBS}
    \text{H}_{BS} =& \frac12 g \mu_b ( \delta B_z \Lambda'_3 - \frac{1}{\sqrt{2}} \delta B_x \Lambda'_1 + \frac{1}{\sqrt{2}} \delta B_x \Lambda'_3 \nonumber\\
    &+ \frac{1}{\sqrt{2}} \delta B_y \Lambda'_2+\frac{1}{\sqrt{2}} \delta B_y \Lambda'_6) \nonumber\\
    =&\frac12 g \mu_b \small\begin{pmatrix}
         0 &  \frac{-1}{\sqrt{2}} (\delta B_x + i\delta B_y) & \delta B_z   & \frac{1}{\sqrt{2}} (\delta B_x - i\delta B_y) \\
         \frac{-1}{\sqrt{2}} (\delta B_x - i\delta B_y) & 0 & 0 & 0 \\
         \delta B_z & 0 & 0 & 0 \\
         \frac{1}{\sqrt{2}} (\delta B_x + i\delta B_y) & 0 & 0 & 0  
    \end{pmatrix}.  
\end{align} 

\noindent And the Hamiltonian of a DQD is:
\begin{widetext}
\begin{align}
    \text{H} &= \text{H}_{\eta} + \begin{pmatrix}
        0 & \\
         & \text{H}_{T_0,T_\pm}
    \end{pmatrix} + \text{H}_{BS} \nonumber \\
    &= \frac12 g \mu_B \begin{pmatrix}
        -\frac{1}{4g \mu_B}  \text{J}_\text{exc} & \frac{-1}{\sqrt{2}} (\delta B_x + i\delta B_y) & \delta B_z  & \frac{1}{\sqrt{2}} (\delta B_x - i\delta B_y) \\
        \frac{-1}{\sqrt{2}} (\delta B_x - i\delta B_y) & \frac{1}{4g \mu_B}  \text{J}_\text{exc} + B_z  & \frac{1}{\sqrt{2}} \left(B_x -i B_y\right) & 0 \\
        \delta B_z & \frac{1}{\sqrt{2}} \left(B_x +i B_y\right) & \frac{1}{4g \mu_B}  \text{J}_\text{exc} & \frac{1}{\sqrt{2}} \left(B_x -i B_y\right) \\
        \frac{1}{\sqrt{2}} (\delta B_x + i\delta B_y) & 0 & \frac{1}{\sqrt{2}} \left(B_x +i B_y\right) & \frac{1}{4g \mu_B}  \text{J}_\text{exc}-B_z
    \end{pmatrix}. \label{EqA: HamilSym}
\end{align}
\end{widetext}

\noindent It is important to note that the sign of the exchange coupling $ \text{J}_\text{exc}$ does not change the dynamics of the DQD system since it has inversion symmetry. Once the Lie algebra and its generators  beyond our quantum system are defined, we are able to study the dynamics of a DQD using rotation operators. Following the discussion in Sec.~\ref{Sec: ST0qubits} and using Eq.~\ref{Eq: Rotation}, we can represent a general rotation as:
\begin{widetext}
    \begin{align} \label{EqA: RotationDim4}
        R(\text{H},\tau_g) &= \mathcal{T} \left\{ \exp\left( \frac{-i}{2} \int_0^{\tau_G} dt \text{H}(t) \right) \right\} \\
        &= \mathcal{T} \biggl\{ \exp\biggl( \frac{-i}{2} \int_0^{\tau_G} dt \biggl[\frac18  \text{J}_\text{exc} \eta + \frac12 g \mu_B ( \frac{B_z}{2}\Lambda_3 + \frac{\sqrt{3}B_z}{2} \Lambda_8 + \frac{1}{\sqrt{2}} B_x \Lambda_1 + \frac{1}{\sqrt{2}} B_x \Lambda_6 + \frac{1}{\sqrt{2}} B_y \Lambda_2 + \frac{1}{\sqrt{2}} B_y \Lambda_7 ) \nonumber  \\ 
         &\hspace{4cm}+ \frac12 g \mu_b ( \delta B_z \Lambda'_3 - \frac{1}{\sqrt{2}} \delta B_x \Lambda'_1 + \frac{1}{\sqrt{2}} \delta B_x \Lambda'_3 + \frac{1}{\sqrt{2}} \delta B_y \Lambda'_2+\frac{1}{\sqrt{2}} \delta B_y \Lambda'_6) \biggr]\biggr) \biggr\},
    \end{align}
    
where the rotation axis $\hat{r}$ is given by:
\begin{align}
    \hat{r } = \frac{H(t)}{\sqrt{\left(\frac18  \text{J}_\text{exc}\right)^2 + \left(\frac12 g \mu_B\right)^2 \left( \left(\frac{B_z}{2}\right)^2 + \left(\frac{\sqrt{3}B_z}{2}\right)^2 + 2\left(\frac{1}{\sqrt{2}} B_x\right)^2 + 2\left(\frac{1}{\sqrt{2}} B_y\right)^2  + \left(\delta B_z\right)^2 + 2\left(\frac{1}{\sqrt{2}} \delta B_x\right)^2+ 2\left(\frac{1}{\sqrt{2}} \delta B_y\right)^2 \right) }}.
\end{align}

\end{widetext}
\section{Mathematical rotations and other encoded spin qubits}\label{SecA: Other encoded qubits}

\subsection{Rotations of a ST$_0$ qubit}
\noindent Mathematically, rotations of a ST$_0$ qubit can be represented as:
\begin{align}
    \hat{\sigma}^x &= S_1^z-S_2^z , \\
    \hat{\sigma}^y &= 2 \hat{z}\cdot \vec{S_2} \times \vec{S_1}, \\
    \hat{\sigma}^z &=2\left(\vec{S_1}\cdot\vec{S_2}\right) - S_1^z S_2^z.
\end{align}

\subsection{Other spin qubits encoded in a 2-electron DQD system}
\subsubsection{Flip-Flop qubits}

\noindent The Flip-Flop qubits are encoded within a DQD system, the computational basis is given by:
\begin{align}
    \ket{0}&= \ket{\uparrow\downarrow}= \frac{1}{\sqrt{2}}\left(\ket{S}+\ket{T_0}\right),\\
    \ket{1}&= \ket{\downarrow\uparrow} = \frac{1}{\sqrt{2}} \left(\ket{S}-\ket{T_0}\right).
\end{align}
In this case, the energy levels are a liner combination of the singlet and T$_0$ triplet.  The rotations are controlled by changing the exchange coupling with the detuning $\varepsilon$ and by the external magnetic fields:
\begin{align}
    \hat{\sigma}^x&= 2\left(\vec{S_1}\cdot\vec{S_2}\right) - S_1^z S_2^z, \\
    \hat{\sigma}^y &= 2 \hat{z}\cdot \vec{S_2} \times \vec{S_1}, \\
    \hat{\sigma}^z &= S_1^z-S_2^z.
\end{align}

\subsection{ST$_\pm$ Qubits}
\noindent The ST$_+$ qubits are encoded within a DQD system, the computational basis is given by:
\begin{align}
    \ket{0}&= \ket{S}, \\
    \ket{1}&= \ket{\uparrow\uparrow} \text{ or } \ket{\downarrow\downarrow}.
\end{align}

\noindent The energy splitting between the computational basis is given by the gradient of the external magnetic field $\Delta B_z$.  The rotations are controlled by the potential of each QD and the potential of the barrier, to tune the exchange coupling, and by the external magnetic field:
\begin{align}
    \hat{\sigma}^x &= \left(S_2^x-S_1^x\right)/\sqrt{2} + \sqrt{2}\hat{y}\cdot \vec{S_2} \times \vec{S_1} \\
    \hat{\sigma}^y &= \left(S_1^y-S_2^y\right)/\sqrt{2} + \sqrt{2}\hat{x}\cdot \vec{S_1} \times \vec{S_2}  \\
    \hat{\sigma}^z &= -\left(S_1^z+S_2^z\right)/2 -\vec{S_1}\cdot\vec{S_2} -S_1^zS_2^z.
\end{align}

\section{Dyson series} \label{SecA: Dyson series}
\begin{align}
    \hat{U}(t,t_0)&\approx1 - \frac{i}{\hbar} \int_{t_0}^t dt_1 \text{H}_I(t) \\
    &+ \left(\frac{-i}{\hbar}\right)^2 \int_{t_0}^t dt_1 \int_{t_0}^{t_1}dt_2 \text{H}_I(t_1)\text{H}_I(t_2),
\end{align}

where $\text{H}_I$ is formed by the off-diagonal terms of $\text{H}$, the interaction terms. The Hamiltonian is given in Eq.~\ref{EqA: HamilSym} and we assume that the pulses have a constant value during the gate:
\begin{align}
    \hat{U}_I(t,t_0)&\approx1 + \frac{-i}{\hbar} \int_{t_0}^t dt_1 \text{H}_I(t) \\ 
    & \hspace{0.4cm}+ \left(\frac{-i}{\hbar}\right)^2 \int_{t_0}^t dt_1 \int_{t_0}^{t_1}dt_2 \text{H}_I(t_1)\text{H}_I(t_2) \\
    &=1 - \frac{i}{\hbar} \text{H}_I t + \frac{1}{2} \left(\frac{-i}{\hbar}\right)^2 (\text{H}_I t )^2.
\end{align}

\noindent If the initial state is $\ket{S}$, the first order in the Dyson series terms would be given by $\bra{\psi}\int_{t_0}^t dt_1 \text{H}_I(t_1)\ket{S}$:
\begin{align}
    A^I_{S \xrightarrow{} S} &= 0, \\
    A^I_{S \xrightarrow{} T_0} &=\frac{1}{2} g \mu_B \delta B_z t, \\
    A^I_{S \xrightarrow{} T_+} &= \frac{1}{2} g \mu_B [\frac{-1}{\sqrt{2}}(\delta B_x - i\delta B_y)] t, \\
    A^I_{S \xrightarrow{} T_-} &= \frac{1}{2} g \mu_B [\frac{1}{\sqrt{2}}(\delta B_x + i\delta B_y)] t.
\end{align}

The second order terms of the dyson series that come back to the state $\ket{T_0}$ are given by $\bra{T_0}\int_{t_0}^t dt_1 \int_{t_0}^{t_1}dt_2 \text{H}_I(t_1)\text{H}_I(t_2)\ket{S}$:

\begin{align}
    A^I_{S \xrightarrow{} S \xrightarrow{} T_0} &= 0, \\
    A^I_{S \xrightarrow{} T_0 \xrightarrow{} T_0} &= 0, \\
    A^I_{S \xrightarrow{} T_+ \xrightarrow{} T_0} &= \frac{-1}{8} (g \mu_B)^2 (\delta B_x - i\delta B_y)( B_x + i B_y)\frac{t^2}{2}, \\
    A^I_{S \xrightarrow{} T_- \xrightarrow{} T_0} &= \frac{1}{8} (g \mu_B)^2 (\delta B_x + i\delta B_y)( B_x -i B_y)\frac{t^2}{2}.
\end{align}

Then the time evolution operator in the interaction picture to go from the state $\ket{S}$ to the state $\ket{T_0}$ up to second order is:
\begin{widetext}
    \begin{align}
        \hat{U}^{S\xrightarrow{}T_0}_I(t,0)&\approx  \frac{-it}{\hbar} \frac{1}{2} g \mu_B \delta B_z   + \frac12\left(\frac{-it}{\hbar}\right)^2 \left(\frac{1}{2} g \mu_B\right)^2 \left( \frac{-1}{2}(\delta B_x - i\delta B_y) \cdot  ( B_x + i B_y) +  \frac{1}{2}(\delta B_x + i\delta B_y) \cdot ( B_x -i B_y)\right) \\
        &=\frac{-it}{\hbar} \frac{1}{2} g \mu_B \delta B_z   -i\left(\frac{t}{\hbar}\right)^2\left(\frac{1}{2 \sqrt{2}} g \mu_B \right)^2 (\delta B_x B_y-\delta B_y B_x) 
    \end{align}

\end{widetext}

We can compute the evolution up to second order from the state $\ket{T_0}$ to $\ket{S}$ in the same way. Since $\bra{S}\text{H}_I(t)\ket{T_0}= \left(\bra{T_0}\text{H}_I^\dagger(t)\ket{S}\right)^\dagger = \left(\bra{T_0}\text{H}_I(t)\ket{S}\right)^\dagger$ the evolution is given by the same expression. The effective amplitude of the transition from the state $\ket{S}$ to $\ket{T_0}$ in the Schrödinger picture is given by:

\begin{align}
    A_{\ket{S} \rightarrow{} \ket{T_0}} &=  A^0_{\ket{S} \rightarrow{} \ket{T_0}} + A^1_{\ket{S}  
    \rightarrow{} \ket{T_0}}  + A^2_{\ket{S} \rightarrow{} \ket{T_0}},
\end{align}    
where
\begin{align}
    A^0_{\ket{S} \rightarrow{} \ket{T_0}} =& 0 \\
    A^1_{\ket{S} \rightarrow{} \ket{T_0}} =& \frac{1}{2} g \mu_B \delta B_z
\end{align}
\begin{widetext}
To compute the second order amplitude, we have to start with the integral amplitude in the interaction picture: 
\begin{align}
     A^{I;2}_{\ket{S} \rightarrow{} \ket{T_0}} = & \sum_{m= T_\pm} 
    \left(\frac{-i}{\hbar} \right)^2 \int_{t_0}^t dt_1 \int_{t_0}^t dt_2 \bra{T_0}H_I(t_1)\ket{m} \bra{m}H_I(t_2)\ket{S},
\end{align}
and rewrite it in the Schrödinger picture:
\begin{align}
    A^{S;2}_{\ket{S} \rightarrow{} \ket{T_0}} = & \sum_{m= T_\pm} 
    \left(\frac{-i}{\hbar} \right)^2 \int_{t_0}^t dt_1 \int_{t_0}^t dt_2 \bra{T_0}H_I(t_1)\ket{m} \bra{m}H_I(t_2)\ket{S} e^{-i(\lambda_{\ket{m}}-\lambda_{\ket{T_0}})t_1/\hbar} e^{-i(\lambda_{\ket{S}}-\lambda_{\ket{m}})t_2/\hbar}.
\end{align}
\end{widetext}
Then, we solve it from $t_0=-\infty$ to $t$:
\begin{align}
    A^{S;2}_{\ket{S} \rightarrow{} \ket{T_0}} = - \sum_{m= T_\pm} \frac{\bra{T_0}H_I\ket{m} \bra{m}H_I\ket{S}}{(\lambda_{\ket{S}}-\lambda_{\ket{m}})(\lambda_{\ket{S}}-\lambda_{\ket{T_0})}}
\end{align}
Finally, the second order transition amplitude from the state $\ket{S}$ to $\ket{T_0}$ is given by:
\begin{align}
    A^2_{\ket{S} \rightarrow{} \ket{T_0}} = \left( \lambda_{\ket{T_0}} - \lambda_{\ket{S}} \right)  A^{S;2}_{\ket{S} \leftrightarrow{} \ket{T_0}} =& \sum_{m= T_\pm} \frac{\bra{T_0}H_I\ket{m}\bra{m}H_I\ket{S}}{\lambda_{\ket{S}} - \lambda_{\ket{m}}}
\end{align} 
Similarly, we can compute the second order transition amplitude from the state $\ket{T_0}$ to $\ket{S}$:
\begin{align}
    A^2_{\ket{T_0} \rightarrow{} \ket{S}} =& \sum_{m= T_\pm} \frac{\bra{S}H_I\ket{m}\bra{m}H_I\ket{T_0}}{\lambda_{\ket{T_0}} - \lambda_{\ket{m}}}
\end{align}

\end{document}